# Theory of coupled electromagnetic circuits, the connection to quantum mechanical resonance interactions and relevance to chronobiology

W. Ulmer[1], Germaine Cornelissen[2], Franz Halberg[2] and Othild Schwarzkopff[2]

[1]Klinikum München-Pasing, Dept. of Radiation Oncology and MPI of Physics, Göttingen, [2]University of Minnesota, Minneapolis, MN, USA

**Abstract**

The existence of specific biorhythms and the role of geomagnetic and/or solar magnetic activities are well-established by appropriate correlations in chronobiology. From a physical viewpoint, there are two different accesses to biorhythms to set up connections to molecular processes: 1. Diffusion of charged molecules in magnetic fields. 2. Quantum mechanical perturbation theoretical methods and their resonance dominators to characterize specific interactions between constituents. The methods of point 2 permit the treatment of molecular processes by circuits with characteristic resonances and 'beat-frequencies', which result from the primarily fast physical processes. As examples the tunneling processes between DNA base pairs (H bonds) and the ATP decomposition are considered.

## 1. Introduction

The description of molecular processes and the energy/charge transport in/between molecules as mechanical (and more promising) electrical oscillators has a long history (Hartmann and Stürmer 1950, Ulmer 1980a. Thus a molecule (or interacting molecules via H bond incorporating an exchange of protons) can be regarded as a certain charge distribution of proper capacitances, whereas certain molecular changes of the configurations are connected by currents. In the early quantum mechanics, Heisenberg used calculations of currents to treat transitions between ground and excited states of atoms to explain spectral properties of them. These transitions usually are very fast processes (the lifetime of excited singlet states amounts to $10^{-7}$ sec; only the lifetime of excited triplet states may varies from $10^{-2}$ sec to minutes and hours). The oscillations of between molecule sites (IR spectra) are much slower (usually a factor $10^{-3}$ – $10^{-4}$ compared to singlet excitations), but still rather fast compared to some biorhythms in cells. It appears that the basic principle of coupled electric oscillators is also useful to study physiological processes for many reasons: It is possible to regard cells as complex systems of charged layers/structures, and all biomolecules are usually highly charged ions (i.e. multipoles). Then it is a consequence to consider cellular systems as various different charge distributions (capacitances) and currents, induced by charge transfer via H bonds or other molecular deformations. This connection indicates that the origin of biochemical resonances is of quantum mechanical nature, since only this tool can determine molecular properties and resonance interactions. In recent time, the development of molecular electronics appears to be a very outstanding devise (Carter 1981, Campbell and Peyrar 1983, Randhawa et al 2011). A further interesting feature is the study of biorhythms. It is one goal of this study to show that biorhythms result from very complicated couplings of electromagnetic oscillators and by splitting of resonance frequencies. By that we man obtain fast oscillations and, in addition, one or two frequencies, which are very slow. In a certain sense this result may be regarded as superimposition of beats to fast oscillator frequencies. Already two coupled electric oscillators are sufficient to study such a model.

## 2. Methods – Theoretical part

### 2.1. Denominations, abbreviations and basic equations

In the following, we make use of the definitions:

*L: inductivitance; C: capacitance; M: mutual inductivitance between two solenoids; U: voltage; Q: electric charge at the capacitance;* $\dot{Q}$ *= dQ/dt: electric current in the solenoid;* $\ddot{Q}$ ***:*** *second time derivative; indices refer to the related oscillator number.*

The basic equation of all electromagnetic processes is the following equation:

$$U_{inductivitance} + U_{capacitance} = 0 \quad (1)$$

A further basic equation is the consideration of one electric oscillator with L and C (Figure 1). From equation (1) and Figure 1 follows:

$$L \cdot \ddot{Q} + Q / C = 0 \quad (2)$$

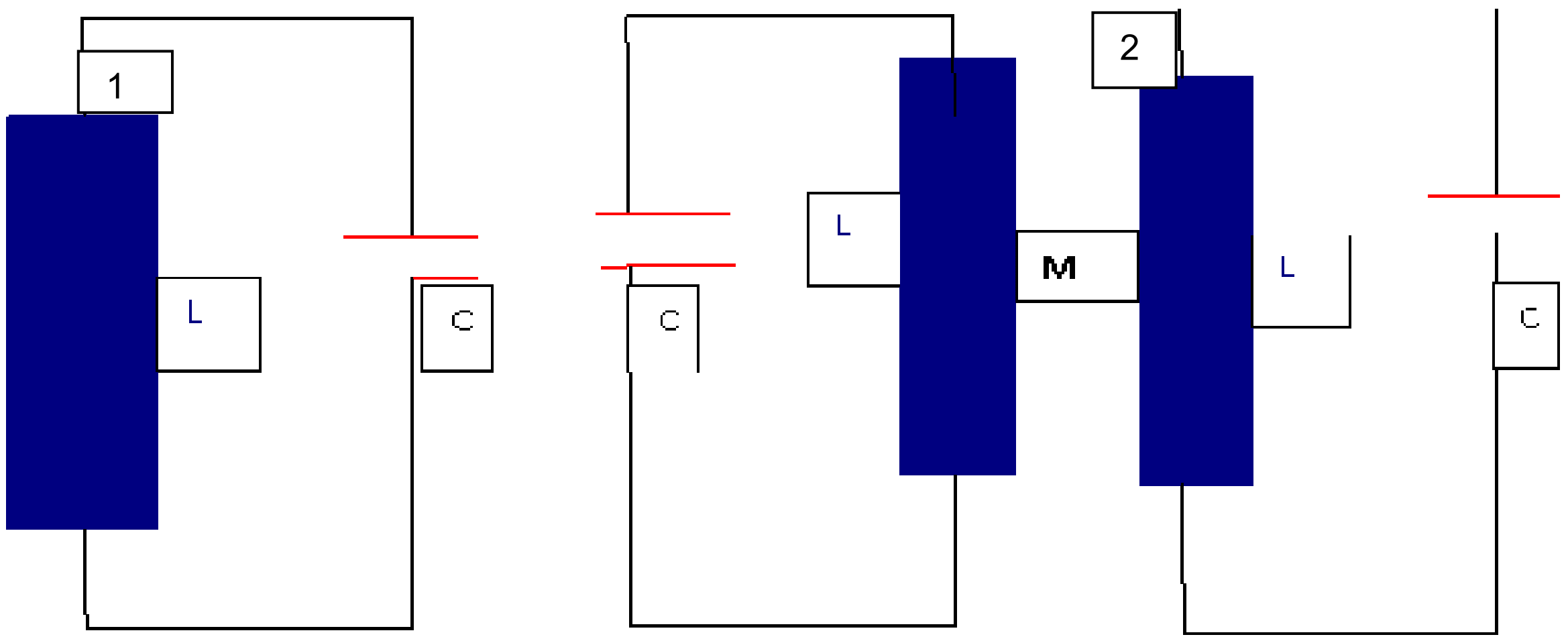

**Figure 1:** One single oscillator (L: inductivitance and C: capacitance).

**Figure 2:** Two identical oscillators according to Figure 1 with mutual coupling M of the currents (magnetic coupling).

The solution of equation (2) is simply given by the 'ansatz':

$$\left.\begin{aligned} Q &= Q_0 \cdot \cos(\omega_0 t) \\ \omega_0^2 &= 1/LC \end{aligned}\right\} \quad (3)$$

It is the task of the following sections to reduce coupled electromagnetic circuits to equation (2) and its solution (3) via the concept of normal modes. Replacing $\cos(\omega_0 t)$ by $\sin(\omega_0 t)$ or by forming either a linear combination of sine and cosine or $\exp(i\ \omega_0 t)$ equation (2) is also satisfied.

### 2.2. Coupling of two/three identical electric oscillators: magnetic coupling via M (coupling constant) and the basic principle of chronobiology

Thus for simplicity we first consider Figures 2 and 3. The basic equations applicable to both Figures are:

$$\left.\begin{aligned} L\ddot{Q}_1 + M(\ddot{Q}_2 + \ddot{Q}_3) + Q_1/C &= 0 \\ L\ddot{Q}_2 + M(\ddot{Q}_1 + \ddot{Q}_3) + Q_2/C &= 0 \\ L\ddot{Q}_3 + M(\ddot{Q}_1 + \ddot{Q}_2) + Q_3/C &= 0 \end{aligned}\right\} \quad (4)$$

Without any restriction equation (4) refers to Figure 3, whereas Figure 2 is described, if the connection of oscillators 1 and 2 to oscillator 3 is quenched by putting M = 0. Then oscillator 3 is completely independent ($Q_3 = Q$) and is treated by equations (2, 3). For this case the normal modes are readily obtained by the substitutions

$$q_1 = Q_1 + Q_2 \text{ and } q_2 = Q_1 - Q_2 \quad (5)$$

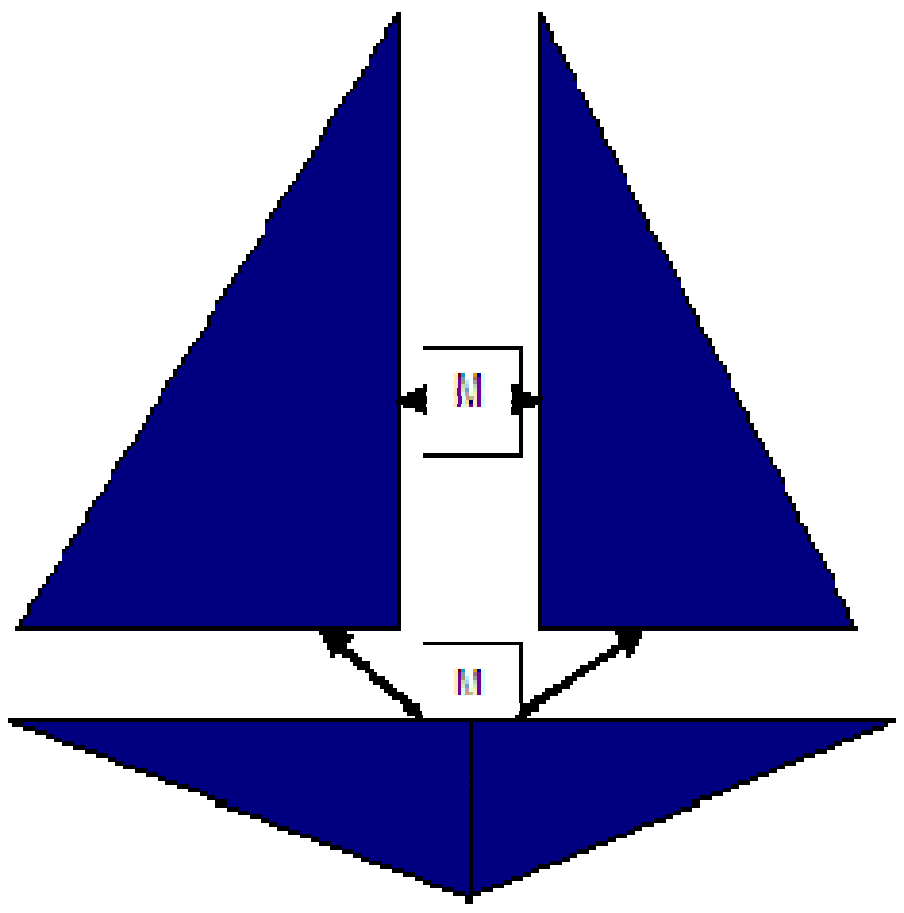

**Figure 3:** Three identical oscillators according to Figures 1 and 2 with coupling between each oscillator: 1 with 2, 2 with 3 and 3 with 1.

The solutions resulting from equation (5) are identical with those of equation (2), if the normal modes $q_1$ and $q_2$ are inserted:

$$q_1 = q_{11}\cos(\omega_1 t) + q_{12}\sin(\omega_1 t) \quad (\omega_1^2 = 1/(L+M)C) \quad (6)$$

$$q_2 = q_{21}\cos(\omega_2 t) + q_{22}\sin(\omega_2 t) \quad (\omega_2^2 = 1/(L-M)C) \quad (7)$$

The arbitrary amplitudes can be fixed by proper initial conditions. By taking $M \rightarrow 0$ the connection between the oscillators is removed and $\omega_1 = \omega_2 = \omega_0$ is valid. The normal modes of 3 coupled oscillators (Figure 3) are obtained by the substitutions:

$$q_1 = Q_1 + Q_2 + Q_3; \; q_2 = Q_1 - Q_2; \; q_3 = Q_1 - Q_3 \quad (8)$$

The solutions in terms of normal modes are:

$$q_1 = q_{11}\cos(\omega_1 t) + q_{12}\sin(\omega_1 t) \quad (\omega_1^2 = 1/(L+2M)C) \quad (9)$$

$$q_2 = q_{21}\cos(\omega_2 t) + q_{22}\sin(\omega_2 t) \quad (\omega_2^2 = 1/(L-M)C) \quad (10)$$

$$q_3 = q_{31}\cos(\omega_3 t) + q_{32}\sin(\omega_3 t) \quad (\omega_3^2 = 1/(L-M)C) \quad (11)$$

Obviously the solutions (9 - 11) agree with the solution (3), if $M \rightarrow 0$ is carried out.

### 2.2.1 Pendular movements - their implications to biorhythms and chronobiology

Due to the coupling M the resonance frequency $\omega_1$ according to equations (6 - 7) and (9 - 11) the frequencies resulting from L+M or L+2M in the denominator can be reduced, whereas according to (7), where the difference L - M enters the denominator; $\omega_2$ may become very high, if M amounts approximately to L. Example: We choose L = C = 1 in such units that $\omega_0$ = 1 and T = 2π days (ca. 6.28 days). Please note that the definition ω = 2πT holds. Then if M = 0.9 L = 0.9 formula (7) provides $T_1 \approx$ ca. 8.5 days and (6) $T_2 \approx 0.3 \cdot 2\pi$, ca. 2 days. However, if M = 0.99 L, then there is no significant change in formula (6), i.e. T ≈ ca. 8.5 days, but formula (7b) provides $T_2$ = 0.6 days. A sudden change of L or C may imply significant changes in the related resonance frequencies. In particular, the denominator of formula (7), where the difference L – M has to be used, may lead to severe changes of the eigenfrequency $\omega_2$ and $T_2$. It follows that the resonance frequencies $\omega_2 = \omega_3$ are degenerate, and only $\omega_1$ is changed; the denominator L + 2M is increased. This provides a decrease of the eigenfrequency $\omega_1$ and a corresponding prolongation of the resonance time $T_1$. Using again the above values C = 1, L = 1 and M = 0.99 L, then $T_1$ amounts to ca. 11 days. In other words: Assume that the unperturbed oscillator shows a circasemi-septan period, then the feed-sideward coupled oscillators (each oscillator couples with 2 other oscillators) may lead to a circaseptan period, if the feed-sideward coupling is strong, as assumed in the above case. Since the solutions (6 - 7) result from two coupled oscillators and both charge amplitudes $q_1$ and $q_2$ linearly depend on $Q_1$ and $Q_2$. We may perform linear combinations of either solution (6) or (7) in order to view the amount of information containing in coupled oscillators. We show this amount in the case of solution (6-7).We make use of a trigonometric theorem and of the substitutions

$$\left.\begin{aligned} &\sin(\alpha+\beta)+\sin(\alpha-\beta)=2\cdot\sin(\alpha)\cdot\cos(\beta)\\ &\cos(\alpha+\beta)+\cos(\alpha-\beta)=2\cdot\cos(\alpha)\cdot\cos(\beta)\\ &\alpha+\beta=\omega_1\cdot t;\quad \alpha-\beta=\omega_2\cdot t \end{aligned}\right\}(12)$$

Then equations (6 - 7) can be rewritten in the form:

$$\left.\begin{aligned} q_{total} =\; & q_{21}\cdot 2\cdot\sin\left(\tfrac{\omega_1 t+\omega_2 t}{2}\right)\cdot\cos\left(\tfrac{\omega_1 t-\omega_2 t}{2}\right)+(q_{22}-q_{21})\cdot\sin(\omega_2 t)\\ & +q_{11}\cdot 2\cdot\cos\left(\tfrac{\omega_1 t+\omega_2 t}{2}\right)\cdot\cos\left(\tfrac{\omega_1 t-\omega_2 t}{2}\right)+(q_{12}-q_{11})\cdot\cos(\omega_2 t) \end{aligned}\right\}(13)$$

Equation (13) provides the information that the superposition amplitude $q_{total}$ contains the basic amplitudes, which are the differences $q_{22} - q_{21}$ and $q_{12} - q_{21}$, resply., and the frequency $\omega_2$, a very fast oscillation frequency $\omega'_1 = (\omega_1 + \omega_2)/2$ in the sine/cosine and a very slow oscillation frequency $\omega'_2 = (\omega_1 - \omega_2)/2$ in the cosine (or $2 \cdot \pi / T'$). If $q_{22} = q_{21}$ and $q_{12} = q_{11}$, the second terms of equation (9) vanishes and the cosine is referred to as carrier amplitude/frequency incorporating 'beats' between the two coupled oscillators. However, if the amplitudes $q_{11}$ and $q_{12}$, $q_{22}$ and $q_{21}$, considerably differ from each other, then beats are still present, but they do not play the main role. It appears that in chronobiology we have to deal with similar situations, where many fast oscillations simultaneously appear besides very slow oscillating components. Since the starting point of equation (13) are two identical oscillators with one coupling *M* between them – this is rather a model than a very realistic case – it is evident that superposition of more complex conditions lead to many fast oscillations and more than one beat amplitude.

### 2.3. Two oscillators with $L_1$, $C_1$, $L_2$, $C_2$, a common dielectric medium ε and the coupling λ(ε)

It appears that Figure 4 represents also significant information with respect to cellular processes, since a dielectric medium between biomolecules (e.g. water) is rather realistic. The presence of any dielectric medium in a capacitance C may also change its magnitude.

The basic equations according to Figure 4 are:

$$\left.\begin{aligned} &\omega_{10}^2 = 1/L_1C_1 ; \; \omega_{20}^2 = 1/L_2C_2 \\ &\ddot{Q}_1 + \omega_{10}^2 Q_1 + (\lambda / L_1) Q_2 = 0 \\ &\ddot{Q}_2 + \omega_{20}^2 Q_2 + (\lambda / L_2) Q_1 = 0 \end{aligned}\right\} (14)$$

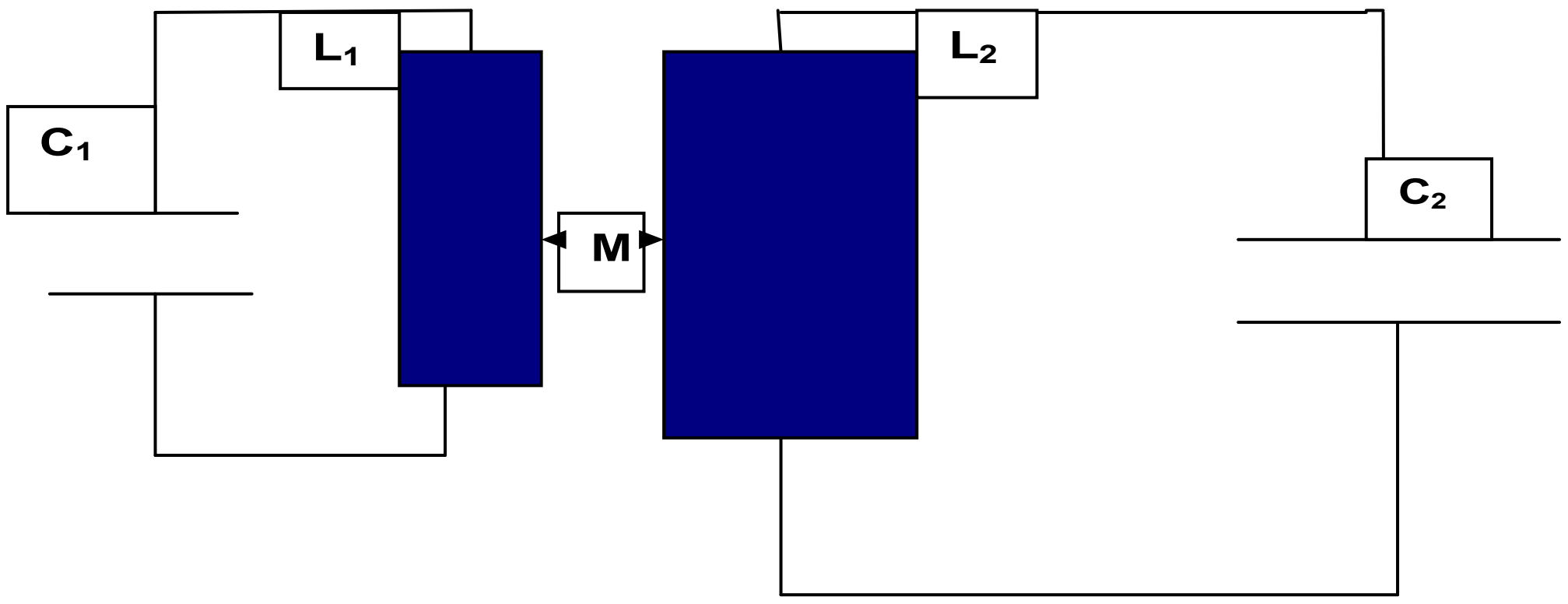

**Figure 4:** Magnetic coupling between 3 different oscillators

The solution procedure is the same as already used:

$$Q_n = Q_{n0} \cdot exp(i\omega t) \quad (n = 1,2) \quad (15)$$

This 'ansatz' provides the matrix equation:

$$\begin{pmatrix} -\omega^2 + \omega_{10}^{\ 2} & \lambda / L_1 \\ \lambda / L_2 & -\omega^2 + \omega_{20}^{\ 2} \end{pmatrix} \cdot \begin{pmatrix} Q_{10} \\ Q_{20} \end{pmatrix} = \begin{pmatrix} 0 \\ 0 \end{pmatrix} \quad (16)$$

The perturbed eigenfrequencies of the electrically coupled oscillators are given by:

$$\omega_{1,2}^{\ 2} = \tfrac{1}{2}(\omega_{10}^{\ 2} + \omega_{20}^{\ 2}) \pm \sqrt{\tfrac{1}{4}(\omega_{10}^{\ 2} + \omega_{20}^{\ 2})^2 - \omega_{10}^{\ 2}\omega_{20}^{\ 2} + \lambda^2 / L_1 L_2} \quad (17)$$

If the connection via a proper dielectric medium is vanishing ($\lambda \rightarrow 0$), we obtain two isolated systems. It should be mentioned that with the help of the frequencies $\omega_1$ and $\omega_2$ we are able to form again two different linear combinations in terms of cosine/sine, and pendular movements according to the solution (13) follow from this behavior. This problem will be discussed in connection with the next section.

### 2.4. Magnetic coupling between 3 different oscillators

The problem is solved similar to equation (4). Instead of unique parameters M, L, C we now consider the case (Figure 5) $L_1$, $C_1$, $L_2$, $C_2$, $M_{12} = M_{21}$, $L_3$, $C_3$, $M_{13} = M_{31}$, $M_{23} = M_{32}$:

$$\left.\begin{aligned} L_1 \ddot{Q}_1 + M_{12} \ddot{Q}_2 + M_{13} \ddot{Q}_3 + Q_1 / C_1 &= 0 \\ L_2 \ddot{Q}_2 + M_{12} \ddot{Q}_1 + M_{23} \ddot{Q}_3 + Q_2 / C_2 &= 0 \\ L_3 \ddot{Q}_3 + M_{13} \ddot{Q}_1 + M_{23} \ddot{Q}_2 ) + Q_3 / C_3 &= 0 \end{aligned}\right\} \quad (18)$$

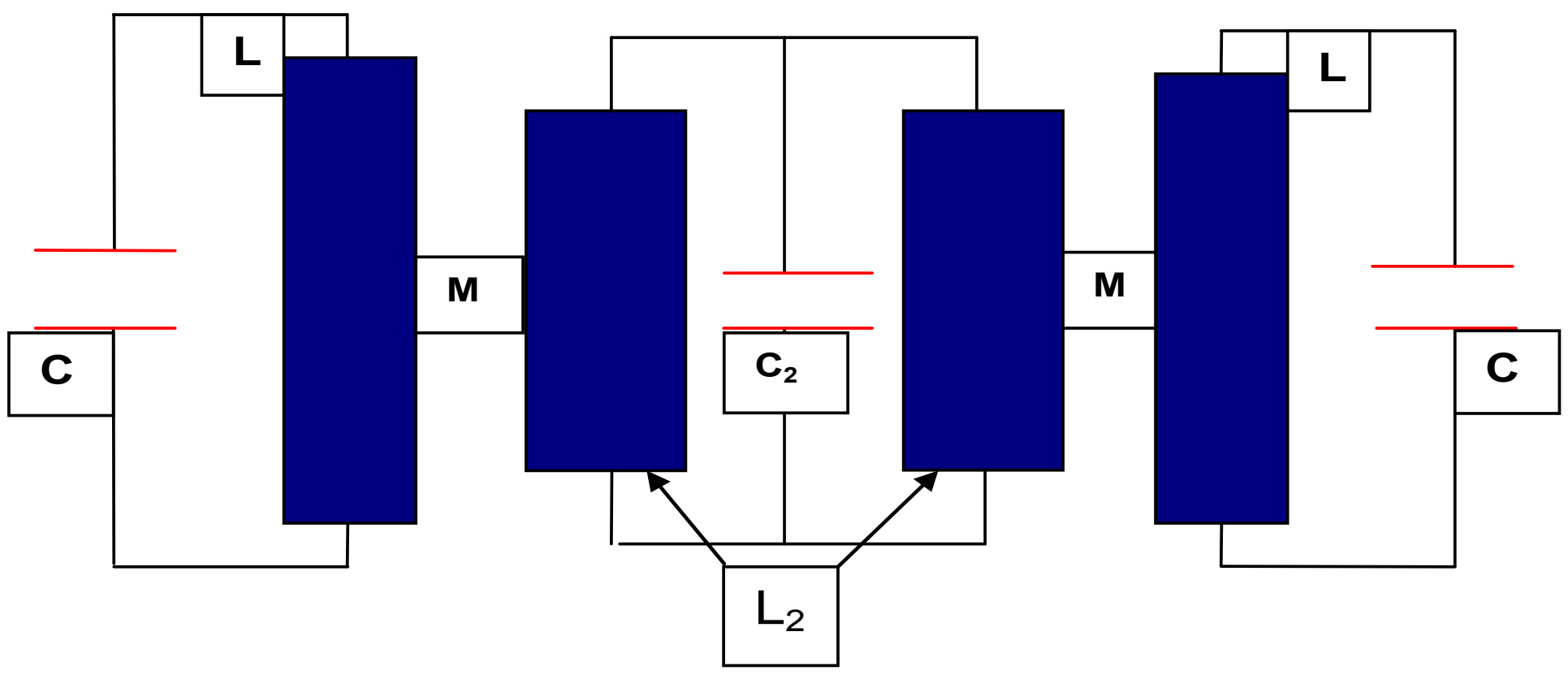

**Figure 5:** Coupled electric oscillators according to equation (37)

The eigenfrequencies of the uncoupled oscillators are:

$$\omega_{10}{}^{2} = 1/L_1C_1 ; \quad \omega_{20}{}^{2} = 1/L_2C_2 ; \; \omega_{30}{}^{2} = 1/L_3C_3 \quad (19)$$

Equation (18) is solved in the same manner as previously used:

$$Q_n = Q_{n,0} \cdot exp(i\omega t) \quad (n = 1,..,3) \; (20)$$

This 'ansatz' requires the solution of the following matrix equation, of which the determinant has to vanish:

$$\begin{pmatrix} -\omega^2 + \omega_{10}{}^2 & -M_{12}\omega^2/L_1 & -M_{13}\omega^2/L_1 \\ -M_{12}\omega^2/L_2 & -\omega^2 + \omega_{20}{}^2 & -M_{23}\omega^2/L_2 \\ -M_{13}\omega^2/L_3 & -M_{23}\omega^2/L_3 & -\omega^2 + \omega_{30}{}^2 \end{pmatrix} \cdot \begin{pmatrix} Q_{1,0} \\ Q_{2,0} \\ Q_{3,0} \end{pmatrix} = \begin{pmatrix} 0 \\ 0 \\ 0 \end{pmatrix} (21)$$

If we perform the substitution $y = \omega^2$ we have to solve the following polynomial equation $y^3 + py^2 + qy + r = 0$. The solution procedure is described in the textbook of Abramowitz and Stegun 1970. All terms resulting from equation (21) are:

$$\left.\begin{aligned}
&y=\omega^2;\; x=y+p/3\\
&div=-1-2M_{12}M_{13}M_{23}/L_1L_2L_3+M_{13}^2/L_1L_3+M_{23}^2/L_2L_3+M_{12}^2/L_1L_2\\
&p=(\omega_{10}^2+\omega_{20}^2+\omega_{30}^2-M_{13}^2\omega_{20}^2/L_1L_3-M_{23}^2\omega_{10}^2/L_2L_3-M_{12}^2\omega_{30}^2/L_1L_2)/div\\
&q=-(\omega_{10}^2\omega_{20}^2+\omega_{10}^2\omega_{30}^2+\omega_{20}^2\omega_{30}^2)/div\\
&r=\omega_{10}^2\omega_{20}^2\omega_{30}^2/div\\
&a=(3q-p^2)/3;\; b=(2p^3-9pq+27r)/27\\
&A=\sqrt[3]{-b/2+\sqrt{0.25b^2+a^3/27}};\; B=\sqrt[3]{-b/2-\sqrt{0.25b^2+a^3/27}}
\end{aligned}\right\}(22)$$

$$\left.\left.\begin{aligned}
&y_1=-p/3+A+B;\\
&y_2=-p/3-(A+B)/2+\tfrac{1}{2}(A-B)\sqrt{-3};\\
&y_3=-p/3-(A+B)/2-\tfrac{1}{2}(A-B)\sqrt{-3};
\end{aligned}\right\}\{\omega^2=y_1,y_2,y_3\}\right\}(23)$$

Remarks: The substitution x = y +p/3 leads the equation $x^3 + ax + b = 0$. The calculation procedure for a, b in terms of p, q, and r are given above in equation (22). If one has calculated A, B in terms of a, b, then the 3 roots are readily computed. With respect to the 3 roots the following cases have to be regarded:

$$\left.\begin{aligned}
&1.\; b^2/4+a^3/27>0\Rightarrow(\text{one real; two conjugate complex})\\
&2.\; b^2/4+a^3/27=0\Rightarrow(3\text{ real; two identical})\\
&3.\; b^2/4+a^3/27<0\Rightarrow(3\text{ real and unequal})
\end{aligned}\right\}(24)$$

The solution of the matrix equation (21) provides 3 normal modes (linear combination of cosine and sine):

$$q_k=[q_{k1}\cdot cos(\omega_k\cdot t)+q_{k2}\,sin(\omega_k\cdot t)]\;(k=1,2,3)\quad(25)$$

It is again possible to construct 'pendular movements' as previously carried out. The manifold increases considerably. We make use of the following definitions:

$$\left.\begin{aligned}
&\omega_{12}=(\omega_1+\omega_2)/2;\; \omega'_{12}=(\omega_1-\omega_2)/2\\
&\omega_{13}=(\omega_1+\omega_3)/2;\; \omega'_{13}=(\omega_1-\omega_3)/2\\
&\omega_{23}=(\omega_2+\omega_3)/2;\; \omega'_{23}=(\omega_2-\omega_3)/2
\end{aligned}\right\}(26)$$

The corresponding overall solution is:

$$\left.\begin{aligned} q_{total} = q_{11} \cdot [&\cos(\omega_{12} \cdot t) \cdot \cos(\omega'_{12} \cdot t) + \cos(\omega_{13} \cdot t) \cdot \cos(\omega'_{13} \cdot t) + \cos(\omega_{23} \cdot t) \cdot \cos(\omega'_{23} \cdot t)] \\ + q_{12} \cdot [&\sin(\omega_{12} \cdot t) \cdot \cos(\omega'_{12} \cdot t) + \sin(\omega_{13} \cdot t) \cdot \cos(\omega'_{13} \cdot t) + \sin(\omega_{23} \cdot t) \cdot \cos(\omega'_{23} \cdot t)] \\ + (q_{21} - q_{11}) &\cdot \cos(\omega_2 \cdot t) + (q_{31} - q_{11}) \cdot \cos\omega_3 \cdot t) \\ + (q_{22} - q_{12}) &\cdot \sin(\omega_2 \cdot t) + (q_{32} - q_{12}) \cdot \sin(\omega_3 \cdot t) \end{aligned}\right\} (27)$$

By that we obtain 3 very slow difference frequencies (‘pendular movements or beats’) and 6 fast (or very fast) frequencies. The 3 different beats may be connected with circadian, circasemi-septan and circaseptan, whereas the fast oscillations might have very short time periods (seconds, minutes or some hours).

## 2.5. Some Generalizations

The preceding section related to Figure 5 may be generalized by two extensions:

1. In addition to the magnetic couplings $M_{12}$, $M_{13}$, and $M_{23}$ it is also possible to introduce connecting dielectric media between the capacitances. Then we have to extend all terms containing $\omega^2$, except the main diagonal elements, in equation (21) by capacitive couplings. The calculation procedure of the eigenfrequencies is the same, only the parameters p, q, r, and consequently a, b, A, B in equation (22) will contain further terms.

2. It is also possible to introduce a fourth oscillator; the coupling to other oscillators may be either magnetic and/or electric, and equation (20) has to be extended to four different charges $Q_1,\ldots,Q_4$:

$$Q_n = Q_{n,0} \cdot \exp(i\omega t) \quad (n = 1,..,4) \quad (28)$$

At every case, we have to solve a polynomial equation of fourth order, if we perform the substitution $y = \omega^2$:

$$y^4 + p \cdot y^3 + q \cdot y^2 + r \cdot y + s = 0 \quad (29)$$

In the textbook of Abramowitz and Stegun 1970 the procedure is described how to find the roots of this equation. A restrictive condition is that $\omega^2 \geq 0$, but negative values have to be excluded. In terms of normal modes the general solution is given by:

$$q_k = q_{k1} \cdot \cos(\omega_k \cdot t) + q_{k2} \cdot \sin(\omega_k \cdot t) \quad (k = 1,...,4) \quad (30)$$

The manifold of 'pendular movements' of the charges between the oscillators is more interesting. On the other side, the difficulties to obtain suitable data also increase. The pendular movements are given by:

$$\left.\begin{aligned} q_{total} &= q_{11}\cdot\sum_{k=1}^{3}\sum_{l=k+1,4}\cos(\omega_{kl}\cdot t)\cdot\cos(\omega'_{kl}\cdot t) + q_{12}\cdot\sum_{k=1}^{3}\sum_{l=k+1,4}\sin(\omega_{kl}\cdot t)\cdot\cos(\omega'_{kl}\cdot t) \\ &+ \sum_{k=2}^{4}(q_{k1}-q_{11})\cdot\cos(\omega_k\cdot t) + \sum_{k=2}^{4}(q_{k2}-q_{12})\cdot\sin(\omega_k\cdot t) \end{aligned}\right\} \quad (31)$$

Figures 6a, 6b are examples of such a generalization.

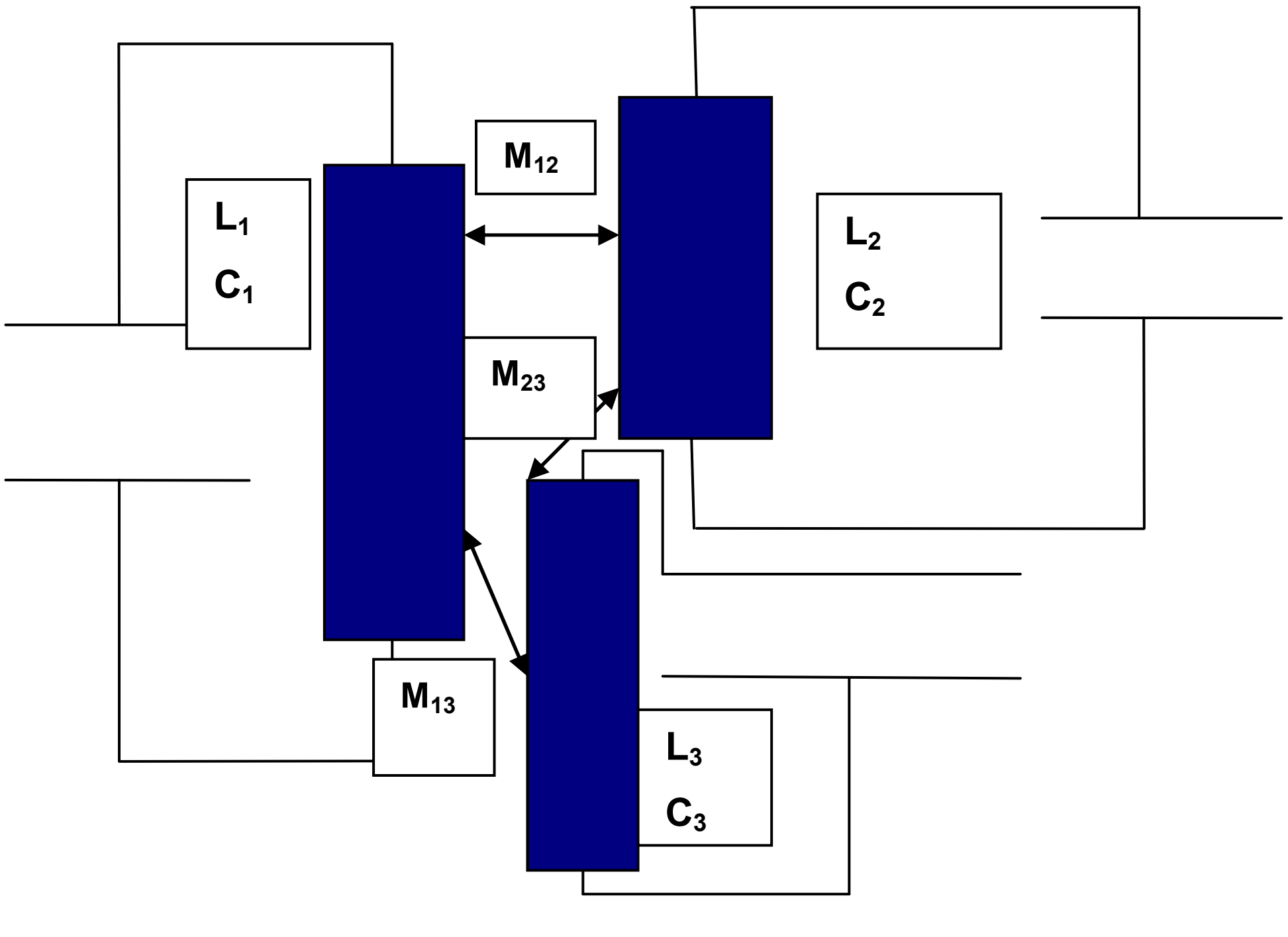

**Figure 6a:** Extension of the coupled circuits to a system with increased complexity compared to Figure 6a.

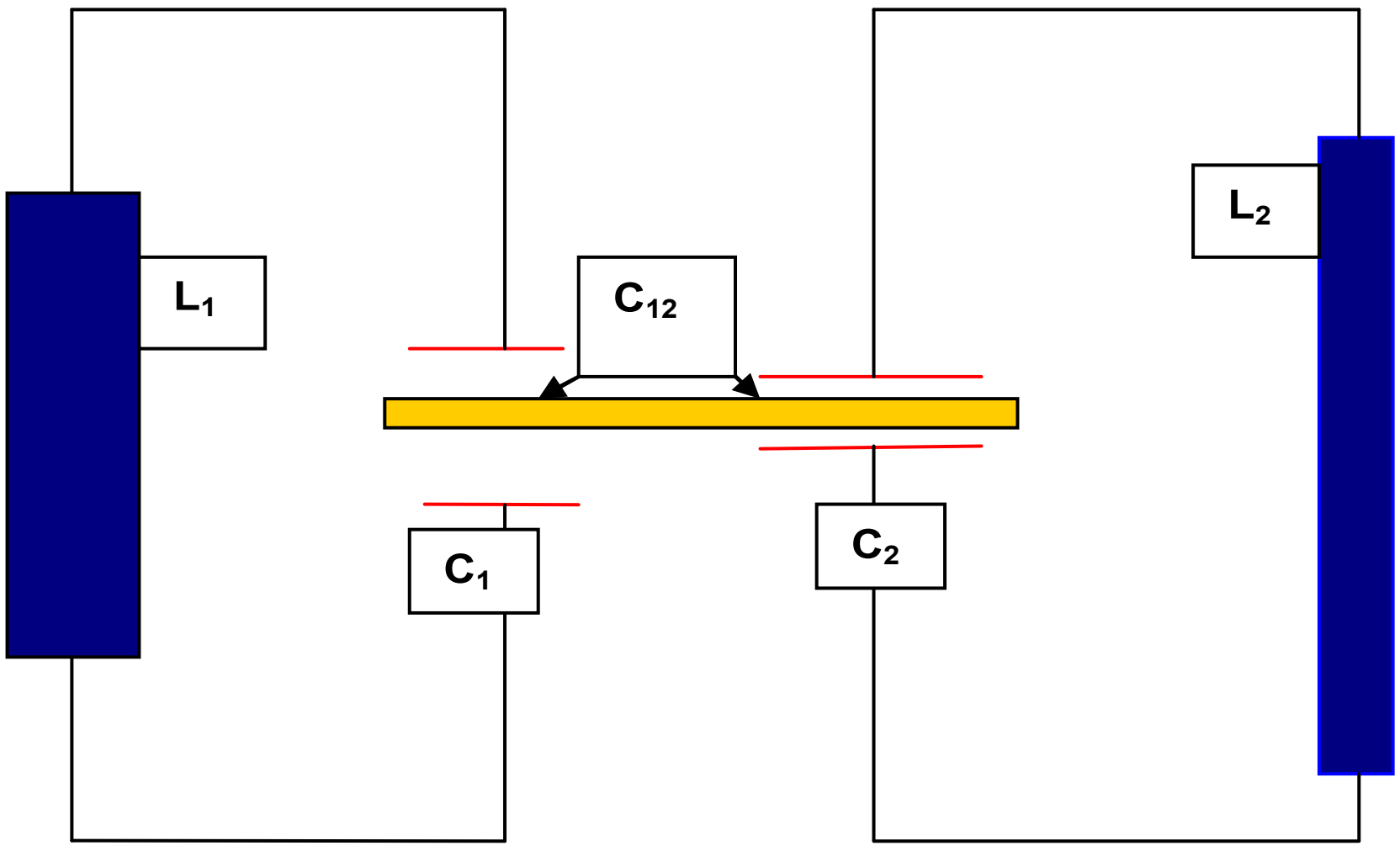

**Figure 6b:** Two different oscillators ($L_1$, $C_1$, $L_2$, $C_2$) with electrical $C_{12}$ coupling by an appropriate dielectric medium.

We should mention that the mathematical treatment of 5 or more coupled oscillators implies numerical evaluations of the corresponding polynomial equations for the eigenfrequencies.

## 2.6. Aspects of pendular movements ('beat frequencies') of coupled circuits

In many fields of applied physics the signal analysis plays a dominant role. In particular, we remember its significance in information and/or control theory and the related technologies. In this case one considers, at least, two coupled electromagnetic circuits; the amplitudes of the normal modes satisfy $q_1 - q_2 \approx 0$ (if more than two coupled circuits are applied, then $q_1 - q_3 \approx 0$, $q_2 - q_3 \approx 0$, etc should also be valid). This situation resembles a mechanical analogue, namely two pendulars with different masses and connected by a spring. In this case $\omega'_{12} = (\omega_1 - \omega_2)/2$ incorporates the extremely slow oscillating ground grequency and $\omega_{12} = (\omega_1 + \omega_2)/2$ is a very fast modulation. The total energy needs s long time to travel from one pendulum to the other one and to return. The pendular movement is therefore referred to as beat frequency ('beats'). The initial conditions have to be chooses such that the very slow ground frequency represents the carrier signal, i.e. $q_1 - q_2 \approx 0$. In macroscopic systems it is always possible to satisfy such initial conditions.

With regard to molecular/cellular biology, where charge distributions of biomolecules (or even cells) are considered as capacitances and charge transfer processes (above all H bonds or interacting metallic ions) as currents, we do not know these initial conditions. This means that fast oscillations might be dominant, and only further components show ‘beats’, i.e. circadian, circasemiseptan and/or circaseptan periods. We have performed an analysis of beat frequencies, i.e. $\omega'_{12} = 2\cdot\pi/\tau_{12}$. It is very astonishing that even for cases where $\omega_1$ and $\omega_2$ refer to fast oscillations ($T_1$ and $T_2$ amount to ca. between 30 seconds and one minute), we are able to obtain numerous circadian, circasemiseptan and circaseptan periods. The only important property is that $\omega_1 \approx \omega_2$ or $T_1 \approx T_2$. Figures 7a – 7d show the yield of circadian, circasemiseptan, circaseptan and circatrigintan (examples have been studied by Halberg et al 1963).

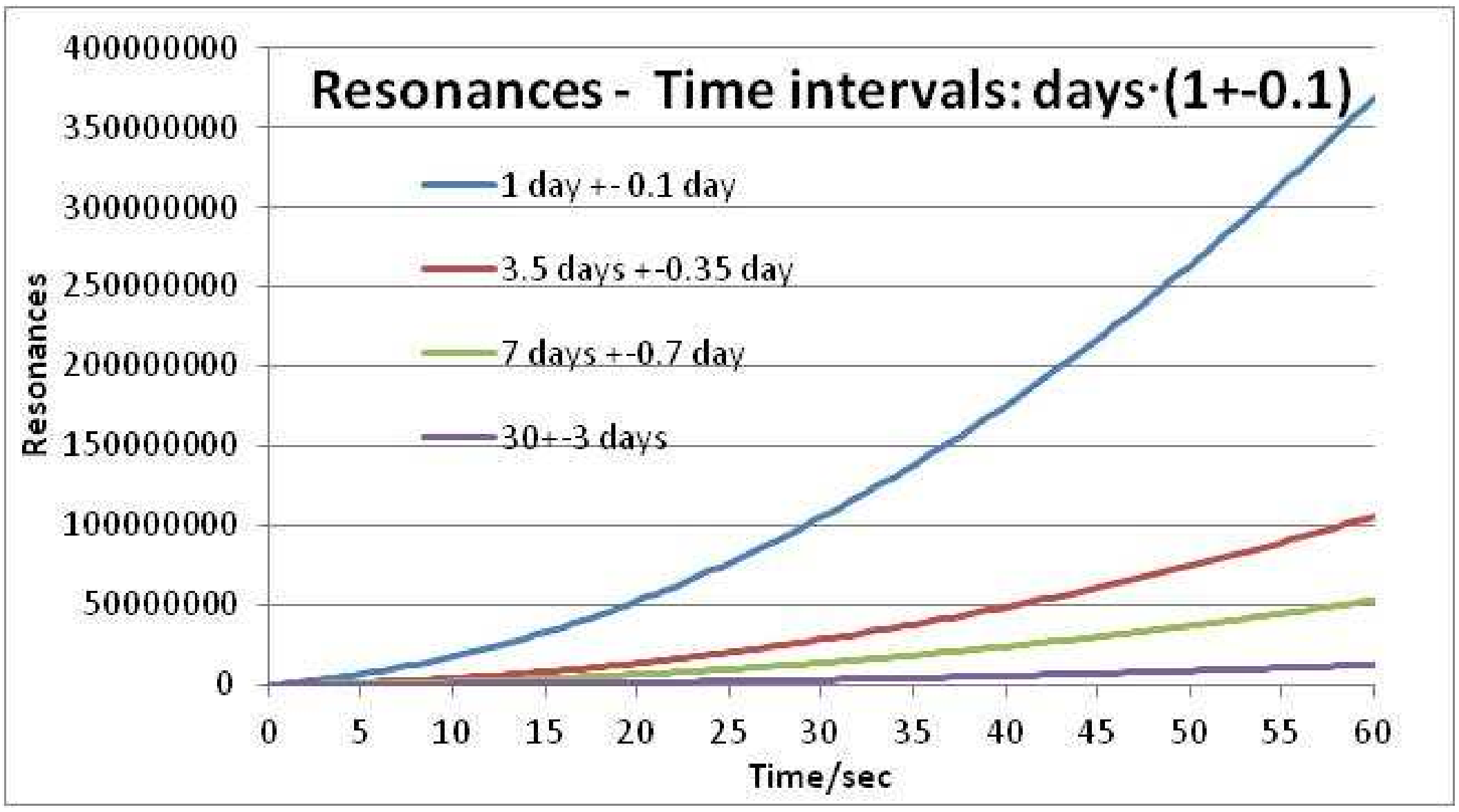

**Figure 7a:** Four coupled oscillator create resonances in the circadian, circa-semiseptan, circaseptan, circatrigintan period. The time raster amounts to ± 10 %.

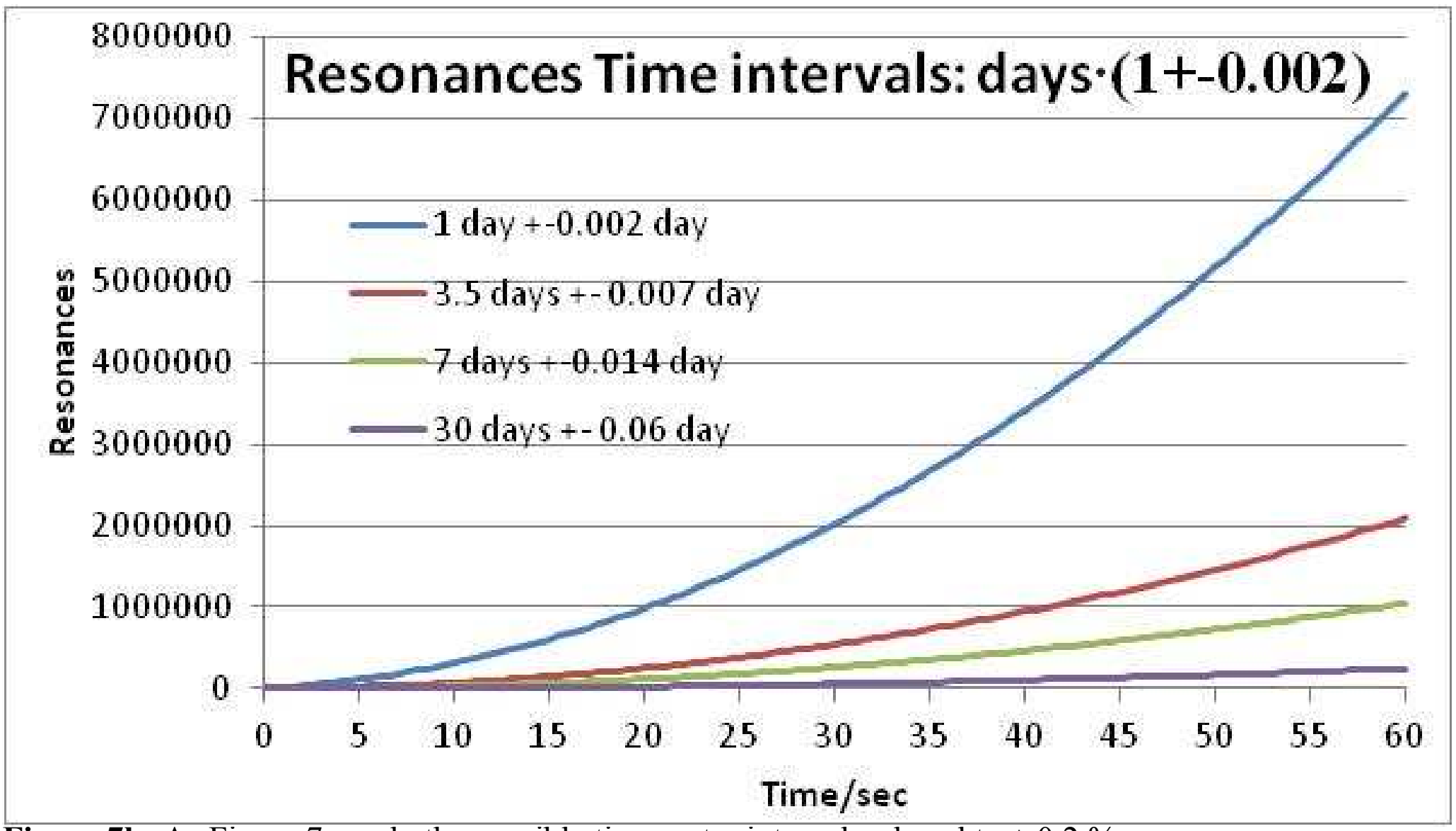

**Figure 7b:** As Figure 7a, only the possible time raster interval reduced to ± 0.2 %.

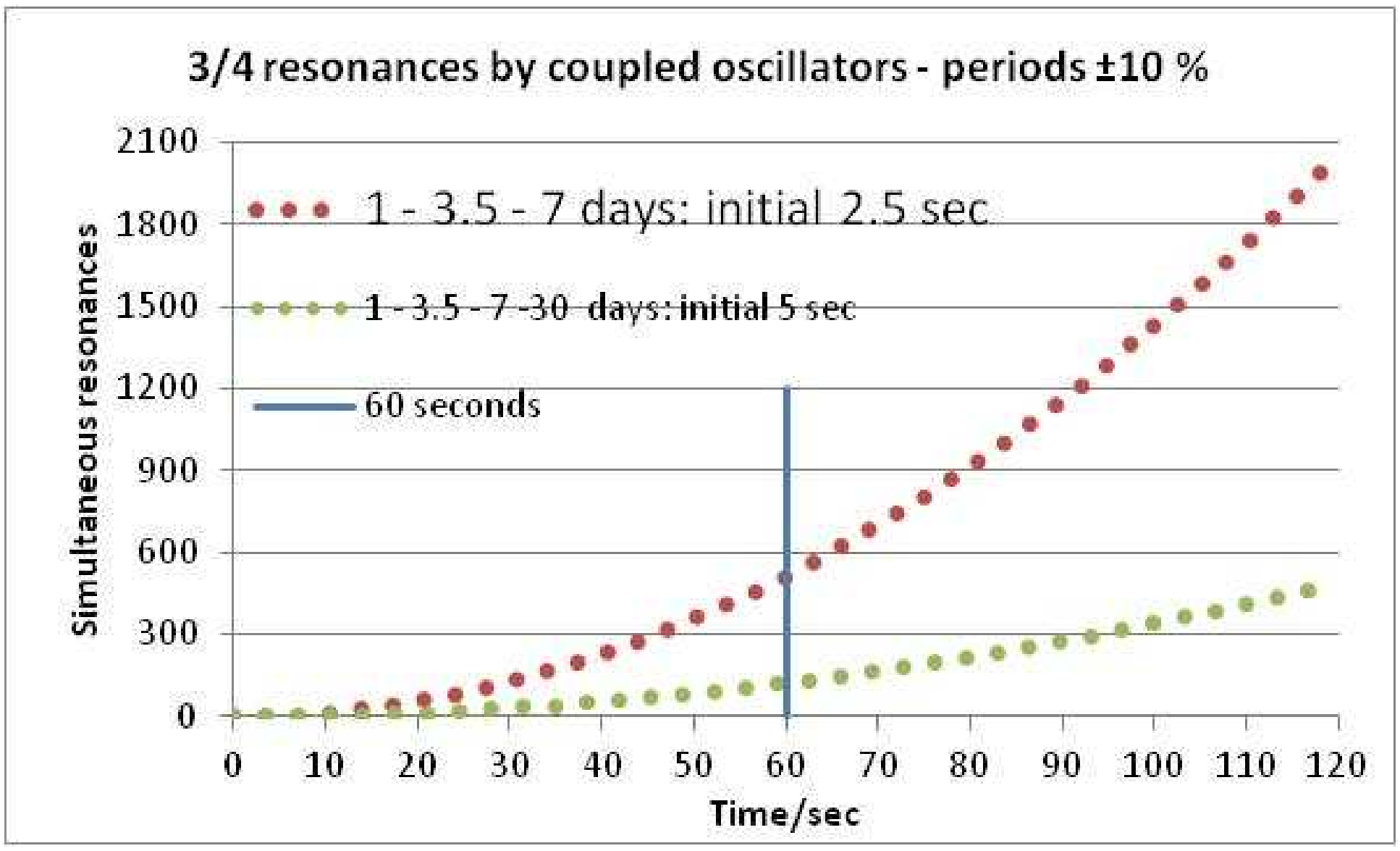

**Figure 7c:** Simultaneous occurrence of resonances of 4 coupled oscillators (3 periods ± 10 %, 4 different periods ± 10 %).

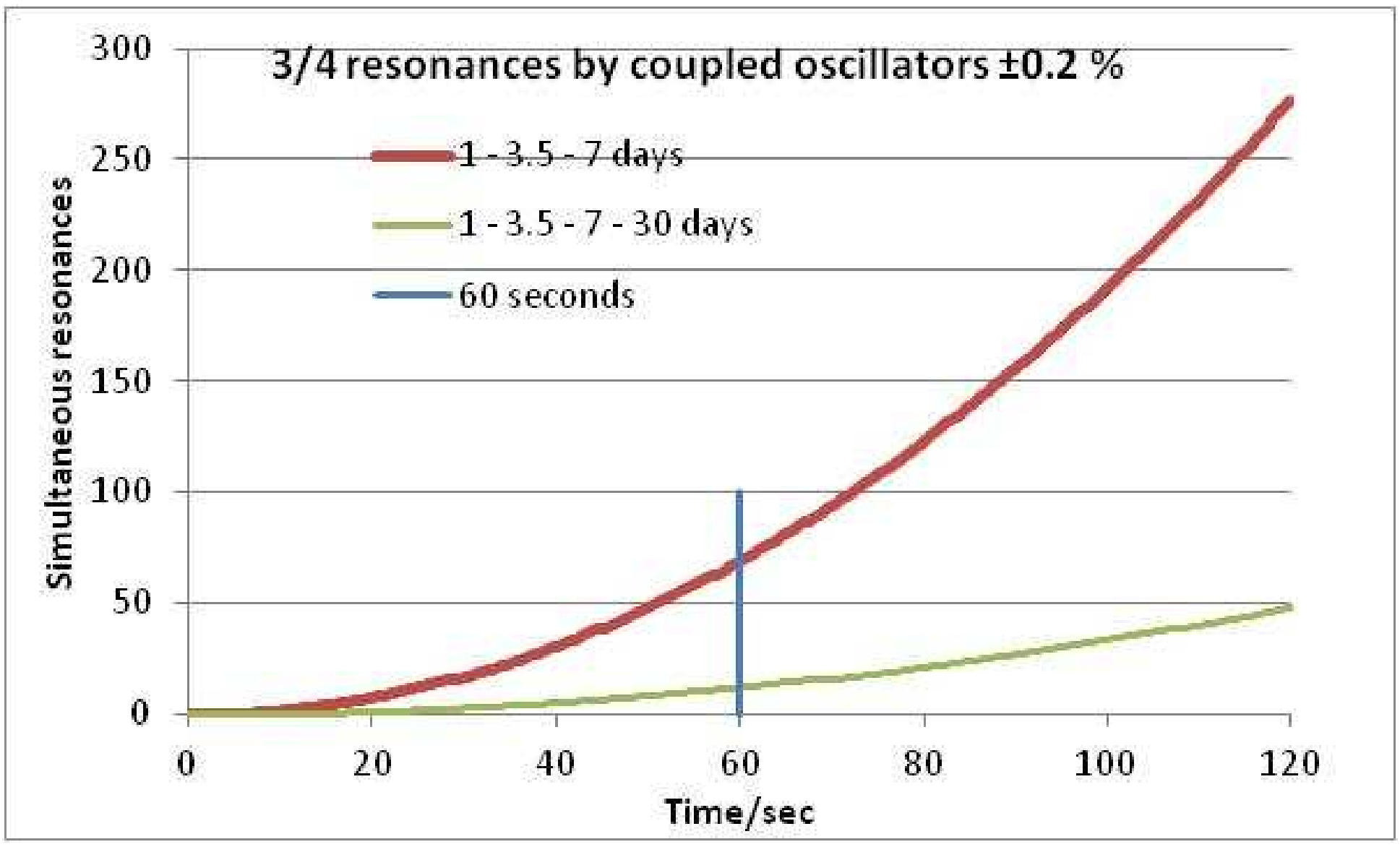

**Figure 7d:** As Figure 7c, the time interval for the occurrence is reduced to 0.2 %.

Figures 7a -7d show the yield of circadian, circa-semiseptan, circaseptan and ca. circatrigintan period $\tau_{12}$ in dependence of $T_1$ and $T_2$. Those cases, where circadian or circasemiseptan, etc are obtained, are significantly increased; therefore we do not show these connections. If we have, at this place, a look to transport phenomena and excitation processes in physics, then the following facts are to verify: Transitions from singlet to triplet states may have life times in the order of seconds to minutes or longer. If excited molecules are coupled or sites of long molecular chains, then the excited state travels due to the proper coupling by inducing molecular deformations and intermediate changes of the electric charge. This is a more or less rather slow process; solitons behave in this manner. Pure singlet excitations imply lifetimes of about $10^{-8}$ seconds. In molecular chains these excitations are referred to as excitons, which are damped by emission of light. Therefore the question arises, what are the principal processes leading to time periods with long durations (biorhythms) studied in chronobiology. Since $\tau_{12}$ may satisfy numerous different kinds of periods, if $\omega'_{12}$ is properly given, the chronobiological findings are not yet founded in satisfactory manner. In other words, we do not yet know the conductors for the preferences of some biorhythms leading to ‘beat time’. One way is to study correlations as performed by the Halberg group (Halberg et al 1965, 1991) and to analyze time series (Schweiger et al 1986). However, we shall return to this question in the following sections.

### 2.7. Periodic oscillators and the transition to the continuum

We continue Figure 2 by introducing further oscillators, where always the direct neighbors are coupled via M. Then we have to analyze the following system of equations ($\omega_0^2 = 1/LC$):

$$\left.\begin{array}{l} L\ddot{Q}_n + M(\ddot{Q}_{n-1} + \ddot{Q}_{n+1}) + Q_n/C = 0 \\ Q_n = q_{n,0}\cdot exp(i\omega t) \end{array}\right\} (32)$$

Equation (31) leads to the following matrix equation

$$\begin{pmatrix} ....0, -M\omega^2/L, -\omega^2+\omega_0^2, -M\omega^2/L, 0...,....., \\ 0..,..0.........., -M\omega^2/L, -\omega^2+\omega_0^2, -M\omega^2/L,..0 \end{pmatrix} \cdot \begin{pmatrix} ..... \\ q_{n-1,0} \\ q_{n,0} \\ q_{n+1,0} \\ ...... \end{pmatrix} = \begin{pmatrix} ... \\ 0 \\ 0 \\ 0 \\ ... \end{pmatrix} (33)$$

From matrix equation (33) follows that the matrix has only diagonal elements of the form $\omega_0^2 - \omega^2$ and directs neighbors with the elements $-M\omega^2/L$. In chapter 9, we also regard the periodic, electric coupling between neighbors. In the continuum limit, this implies a wave equation. With regard to the magnetic coupling according to equation (34), such a wave equation cannot be obtained. On the other side, periodic oscillators with magnetic coupling might have a restricted biological importance, since the periodicity is lacking in such systems.

It is interesting to note that for systems with periodic electric coupling we can derive a wave equation, similar to that considered in continuum physics obtained by periodic mechanical coupling. The analogue to Figure 4 is in the case of electric coupling:

$$L\partial^2 Q_n/\partial t^2 - \lambda(Q_{n+1} + Q_{n-1} - 2\cdot Q_n)/C = 0 \quad (34)$$

In the continuum limit, i.e. the distance $\Delta x$ between the charges $Q_{n+1}$, $Q_n$ and $Q_{n-1}$ becomes infinitely small, equation (34) assumes the shape:

$$\left.\begin{array}{l} L\partial^2 Q_n/\partial t^2 - \lambda\Delta x^2[(Q_{n+1} + Q_{n-1} - 2\cdot Q_n)/\Delta x^2]/C = 0 \\ \lim \Delta x \Rightarrow 0 \end{array}\right\} (35)$$

The term $(Q_{n+1} + Q_{n-1} - 2Q_n)/\Delta x^2$ represents the second derivative with regard to space, if lim $\Delta x \rightarrow 0$ is carried out. Then we obtain a wave equation in the charge space:

$$(\partial^2/\partial t^2 - v^2\partial^2/\partial x^2)\cdot Q = 0 \quad (36)$$

It should be mentioned that in the vacuum case the velocity v is equal to the velocity of light c. But v may be considerably smaller in the presence of a dielectric medium ε, i.e. v ≤ c. With regard to the presumption of equations (32 – 36), namely a periodic system, the same comments are valid as previously pointed out: In biological systems the assumption of periodic boundary conditions may be rather idealistic in contrast to physics of crystals, where polarization waves of molecular crystals are studied. Therefore the study of some coupled oscillators may be more helpful. The inclusion of Ohm's resistance R to be aware of damping (damped waves) is straightforward, and equation (36) assumes the shape:

$$(\partial^2/\partial t^2 + \tfrac{R}{L}\cdot\partial/\partial t - \tfrac{1}{v^2}\partial^2/\partial x^2)\cdot Q = 0 \qquad (37)$$

The mechanical analogue of equation (36) is a system of coupled oscillators with mass m and the force constant f. The equation of motion reads:

$$md^2q_n/dt^2 - f(q_{n+1} + q_{n-1} - 2q_n) \ (38)$$

By introducing the continuum limit in the fashion as in equation (36) we obtain the wave equation of a string with regard to the amplitude (or elongation) q, i.e.

$$(\partial^2/\partial t^2 - (1/v^2)\partial^2/\partial x^2)\cdot q = 0 \ (39)$$

In this case, v is the velocity of sound in the string. In contrast to electrical couplings, the string (solutions of equation (39)) may satisfy well-known constraints, e.g. nodes (q = 0) at x = 0 and x = a. This constraint provides a discrete spectrum of modes:

$$\left.n\cdot\lambda/2 = a; \quad (n = 1,2,...)\right\} \quad (40)$$

In equation (40) λ refers to the wavelength and '*a*' to the distance of the string between the two nodes with q(0) = 0 and q(a) = 0.

The continuum transition of coupled electromagnetic oscillators with M ≠ 0 according to equations (32 - 33) can readily be carried out, but we obtain a 'generalized' wave equation:

$$\left.\begin{aligned} &L\ddot{Q}_n + M(\ddot{Q}_{n+1} + \ddot{Q}_{n-1}) + Q_n/C = 0 \\ &= (L + 2\cdot M)\ddot{Q}_n + M(\ddot{Q}_{n+1} + \ddot{Q}_{n-1} - 2\cdot\ddot{Q}_n) + Q_n/C = 0 \end{aligned}\right\} \ (41)$$

In the continuum limit, we obtain from equation (41)

$$\left.\begin{aligned} &\partial^2/\partial t^2 [1 + l_c^2 \cdot \partial^2/\partial x^2] \cdot Q + \omega^2 Q = 0 \\ &l_c^2 = \Delta x^2 \cdot M/(L + 2M); \quad \omega^2 = \frac{1}{C(L+2M)} \end{aligned}\right\} (42)$$

It is possible to express $\omega^2$ in terms of $\omega_0^2 = 1/LC$, which provides:

$$\omega^2 = \omega_0^2 /(1 + 2\omega_0^2 \cdot M \cdot C) \quad (43)$$

Equation (42) is readily be solved by plane waves (Fourier expansion). Writing the k-vector in the form $k^2 = n^2/l_c^2$ $(n \leq 0)$ we obtain:

$$\left.\begin{aligned} &\omega_n^2 = \omega^2/(n^2 + 1) \\ &\omega_n^2 = \omega_0^2 /[(n^2 + 1) \cdot (1 + 2\omega_0^2 \cdot M \cdot C)] \end{aligned}\right\} (44)$$

A consequence of the solution (44) is that, in spite of $\omega_0^2 = 1/LC$ might imply a fast oscillation of a single circuit, the coupling between the oscillators (chain with $M \neq 0$) must lead to very slow oscillation frequencies $\omega_n^2$ due to formula (51). This is in particular true for $n \gg 1$. The superposition of a ground wave with very slow periodicity and a faster modulation is also possible. This aspect might be interesting in chronobiology.

## 2.8. Standing waves

The question arises, in which way standing wave solutions may become reasonable. For this purpose, we assume that the magnetic coupling M is interrupted at positions $x = 0$ and $x = a$. Then a proper wave mode excited within this interval cannot propagate into the domains $x < 0$ and $x > a$, and analogous boundary conditions hold as for the string. These boundary conditions can be satisfied for pure sine modes, i.e.

$$\left.\begin{aligned} &Q = Q_0 \cdot \exp(i \cdot \omega_n t) \cdot \sin(k \cdot x) \\ &k \cdot a = m \cdot \pi \Rightarrow k = m \cdot \pi / a \quad (m = 1, 2, \ldots) \end{aligned}\right\} (45)$$

Combining this result with equation (44), we obtain

$$\left.\begin{aligned} &n^2 = m^2 \pi^2 l_c^2 / a^2 \quad (m = 1, 2, \ldots) \\ &\omega_m^2 = \omega_0^2 /[(m^2 \pi^2 l_c^2 / a^2 + 1) \cdot (1 + 2\omega_0^2 \cdot M \cdot C)] \end{aligned}\right\} (46)$$

Due to the linearity of equation (42) the superposition of the modes yields the following solution:

$$Q(x,t) = \sum_m Q_{m,0} \cdot \sin(\pi \cdot m \cdot x/a) \cdot \cos(\omega_m \cdot t) \quad (47)$$

The question arises, where in biology such a solution is applicable. Standing waves resulting from magnetic coupling result, when at the 'end-points' x = 0 and x = a, the coupling via M is interrupted by electrically neutral molecules. Then the excited wave amplitudes have to remain within this interval, and they form modes with the ground frequency (very slow) and faster modulations. Such a stationary state may breakdown, if at one endpoint a further molecule changes the nodal conditions and the coupling to the neighboring environment is established. The wave then escapes to reach a path or domain, which could not be reached before due to the lacking coupling interaction. This might be a possible mechanism and effect of neurotransmitters at a certain synapse. The escaped standing wave may be replaced by a new one, when the corresponding condition is reestablished and the wave is excited by further signals.

## 2.9. Theoretical considerations of resonance interactions between biomolecules and the quantum mechanical base

The question arises in which way we can transfer quantum mechanical principles and results to problems of circuits, which represent charge distributions and currents influenced by magnetic fields.

### 2.9.1. Some basic aspects of phenomenological correlations and molecular properties

It might appear that quantum theory cannot provide any information on the problem under consideration, namely resonances interacting molecules. According to a previous study (Ulmer, 1980b) the interaction between molecules can describe the chemical affinity, which is phenomenological described by the Arrhenius equation, by the consideration of suitable term schemes and transition probabilities. This theory includes besides the specific affinity the transition to excited triplet states by visible light or by interaction between proper molecules, which can be characterized by their long life-time and superimposition of transport phenomena similar to diffusion processes. Moreover, the inclusion of magnetic fields and current is quite natural and based on a more fundamental theory as diffusion. The resonances can be viewed in light of circuits as previously considered. Although the previous study was mainly restricted to resonance interaction between drugs and DNA in order to explain very different correlations between cancerogenic molecules and mutagenicity it can readily extended to other aspects such as quasi-degenerate triplet resonances of coupled oscillators, which are the main interest in this investigation. Therefore the question arises, what is connection between chemical affinity of biomolecules and their term schemes with biorhythms.

At first, we consider the classification of drugs with regard to the mutagenicity and, by that, to the carcinogenic effectivity. This problem has a long history, and the polycyclic aromatic hydrocarbons (PAHC) have been investigated by many authors with the help of quantum chemical means. Based on calculations of reactivity indices of ground state properties A. and B. Pullman (1962) have developed the famous K- and L- region model to explain and predict the carcinogenic activity of some PAHCs. However, by inclusion of more available PAHCs than the Pullmans had originally available numerous exceptions could be verified. Therefore this model did not contain the whole truth with regard to the connection between chemical structure/reactivity and carcinogenicity, and further quite different models have been proposed. We particularly mention the correlation of low excited states of PAHCs (Mason 1958), charge transfer mechanisms (resonances) of PAHCs and DNA (Hoffman and Ladik 1962, and the connection between the dipole-dipole resonance interaction of PAHCs and excited states of the amino acid tryptophan (Birks 1970) and the double resonance states of PAHCs with absorption of biophotons (Popp et al 1979). Since each of the mentioned correlations only bears exceptions, Sung (1977) has performed a multiregression analysis in order to bring more light in this shortcoming of methodology.

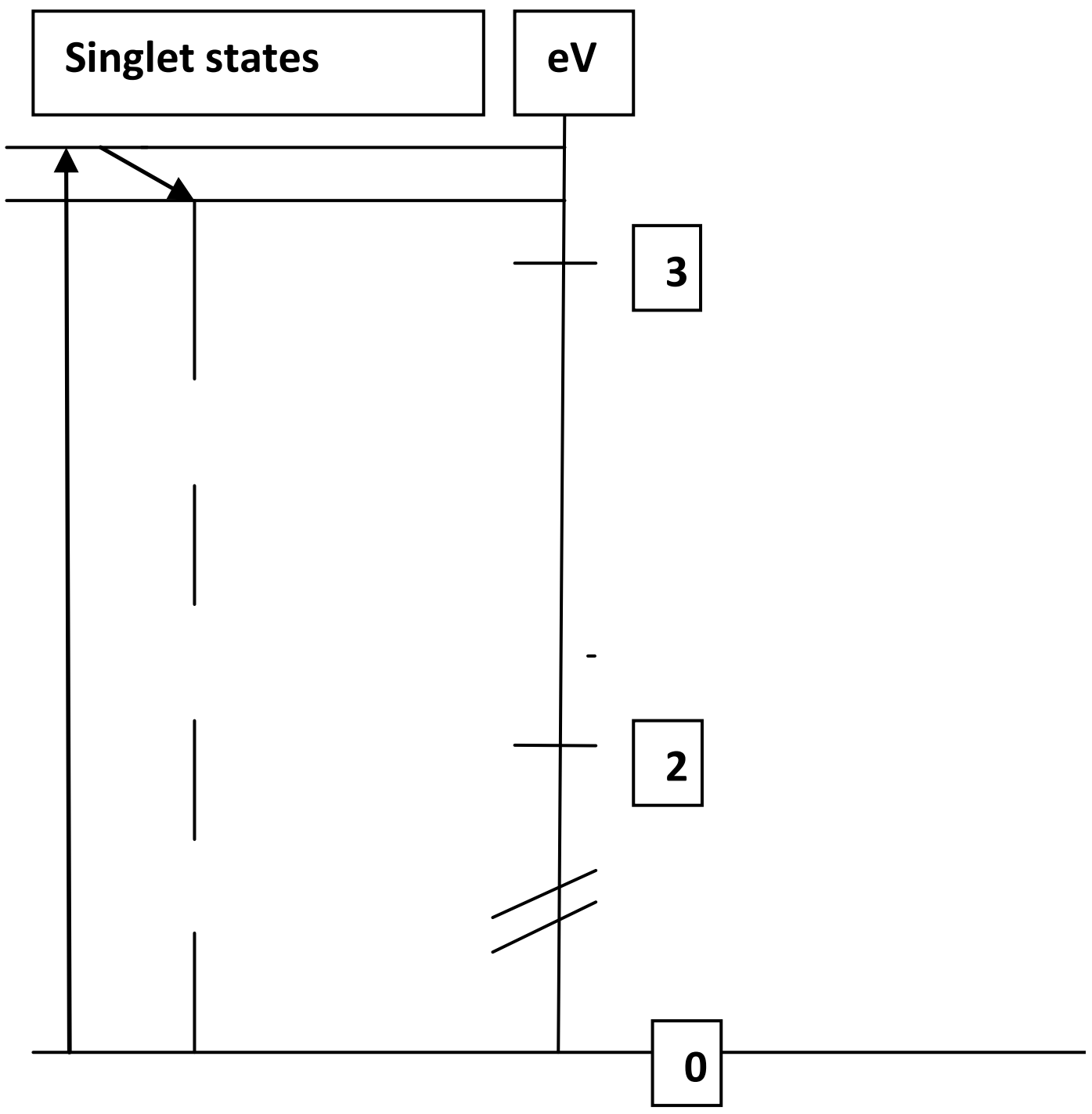

**Figure 8a:** Term scheme of double resonances of carcinogenic PAHCs. The solid arrow represents a permitted transition from the ground state to a higher excited state, whereas the dashes refer to a forbidden transition of the lower excited state to the ground state (the lifetime is significantly increased). The energy difference between these two excited states amounts to ca. 0.1 eV (Popp et al 1979).

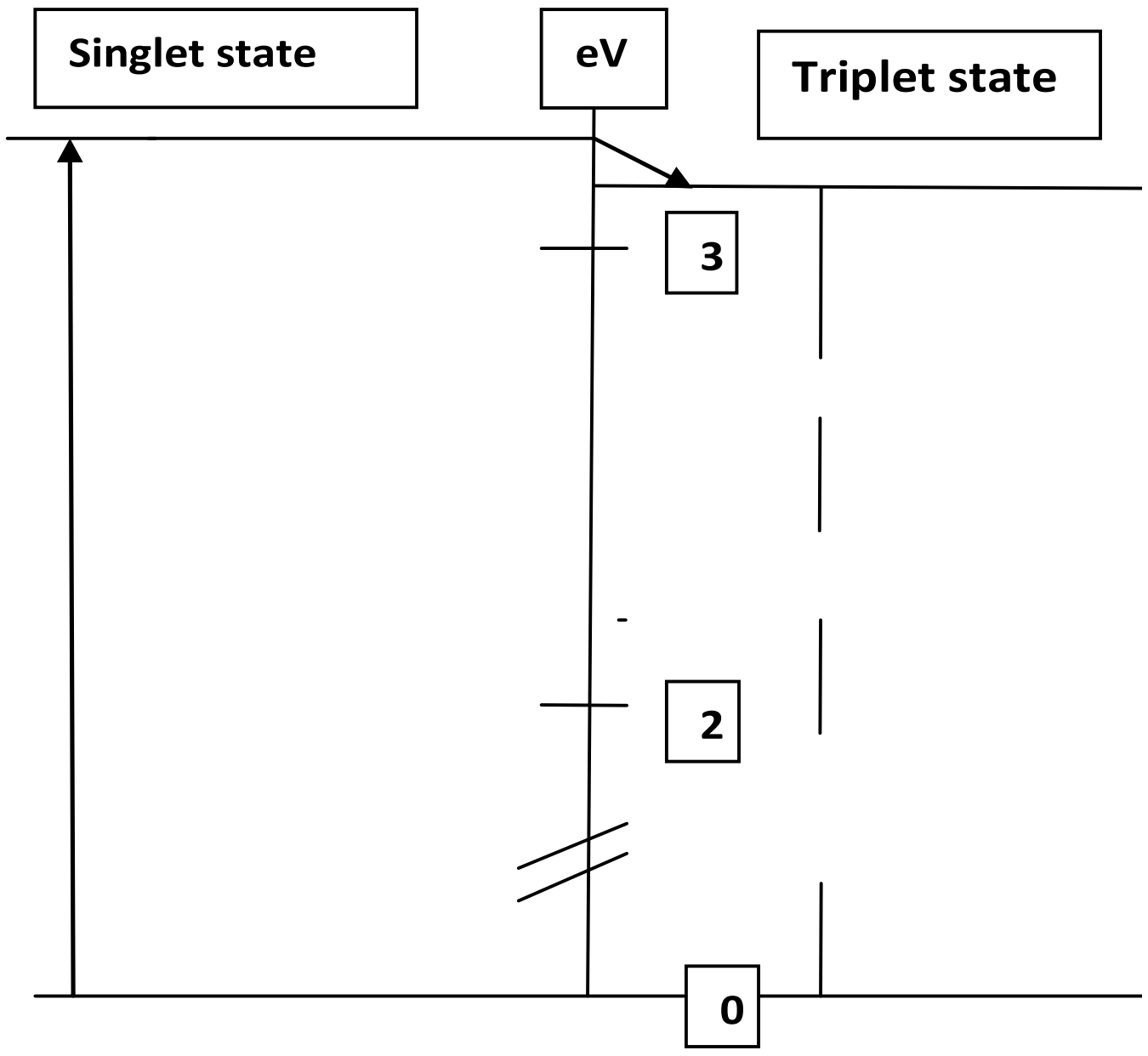

**Figure 8b:** This term scheme incorporates the same effect as Figure 8a, the excitation of the excited triplet state from ground state is forbidden. Spin-orbit coupling permits radiationless transitions from the excited singlet state to the quasi-degenerate triplet state. The lifetime of this state may be very long (order of seconds or minutes).

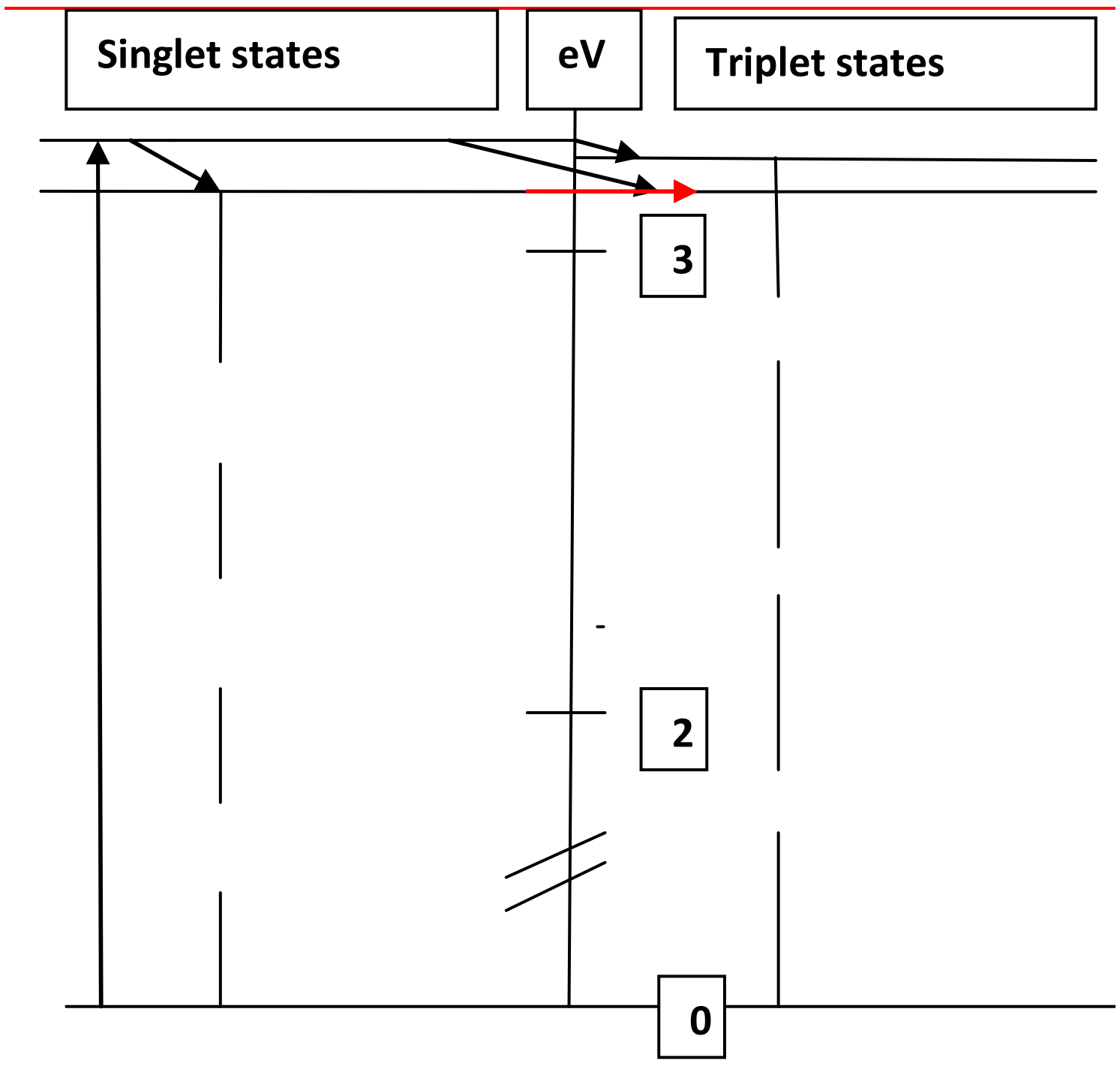

**Figure 8c:** Term scheme of the cytostatic drug cyclophosphamide and its metabolites (Ulmer 1979, 1981). The energy levels of the metabolites are only slightly changed, but the transition probability can drastically differ due to changes of spin-orbit coupling.

According to Ulmer (1980b, 1981) Figures 8a - 8c represent term schemes, which are characteristic for those molecules with a specific interaction with DNA or the amino acid tryptophan; the original restriction to PAHCs is superfluous. With the help of the transition probabilities of the related molecule sites the kind of interaction can be specified (chemical reaction, dipole-dipole resonance interaction, charge transfer, etc.). It should also be pointed out that the term schemes under consideration cannot be restricted to the original molecules, but the metabolites (produced by epoxydation, hydroxylation, carboxylation, etc.) usually show the same term scheme with changed transition probabilities. Thus molecules characterized by the term scheme 8c the chemical reaction mechanism can be classified by biradicalic reactions (Hartmann 1964, Fukui et al 1969).

If for any mutagenic substrate or drug the base pairs of DNA or RNA incorporate the bioreceptors, one of the above term schemes should be applicable, and the ionization energy should also be comparable with the bioreceptor, i.e. ca. 6.9 eV – 7.7 eV. However, it is evident, that the presented findings may also be applied to other resonance interaction, e.g. with protein, hormones, etc. The term scheme may either be similar as it is the case for tryptophan and derivatives or rather different, if bioreceptors with other specific properties have to be accounted for. In every case, the chemical affinity between molecules according to the Arrhenius equations stands in close connection to resonance interactions derived by quantum mechanics. This is the subject of the next chapter.

### 2.9.2. Quantum mechanical aspects and perturbation theory

According to the methodology of quantum mechanics/quantum chemistry there are two main approaches to describe the interaction between any two molecules A and B:

1. We assume the Hamiltonian H of the total system is of the form

$$H = H_A + H_B + H_{AB} \;) \qquad \textbf{(48)}$$

For the sake of simplicity, we suppose that the total system as well as the subsystems A and B shall remain in the singlet ground state. Then we have to calculate (e.g. Hartree-Fock, density functional, Feynman propagators or semi-empirical methods) the ground state energies of the subsystems A and B and of the total system in dependence on all nuclear co-ordinates (in realistic calculations restricted). The Hamiltonian H is assumed as usual (Coulomb interactions between electrons, nuclei, and between electrons among themselves). Magnetic interactions of charged particles, spin-spin and spin-orbit couplings are treated as perturbations and will be introduced separately. Although the total system shall remain in the singlet ground state, corrections by excited contributions via CI methods become then very essential, when A and B undergo interactions, since both molecules have to be stronger deformed and distorted. This fact is already true for separated molecules without the interaction term $H_{AB}$, and the validity of the non-crossing rule is an indication for the relevance of excited configurations.

2. We assume that the Schrödinger equation for $H_A$ and $H_B$ is exactly/approximately solved:

$$\left.\begin{aligned} H_A \cdot \psi_A &= E_A \cdot \psi_A \\ H_B \cdot \psi_B &= E_B \cdot \psi_B \end{aligned}\right\} \qquad \textbf{(49)}$$

For the following considerations the impossibility of the first case is not relevant, as we can *measure and classify* the eigenstates and transition probabilities between the states under various conditions. Now we expand the eigenfunctions of H according to equation (x1) in terms of the eigenfunctions of $H_A$ and $H_B$:

$$\psi = \sum_{k=0}^{\infty} C^{A,B}{}_k \cdot \psi_k{}^{A,B} \quad \textbf{(50)}$$

We make use of the perturbation theory to classify the degree of the approximation (the applicability of the usual perturbation theory is assumed, since the states are not degenerate like isolate H atoms). The first order approximation is determined by the well-known relations for the coefficients:

$$C_{1,k} = \sum_{A,B}\sum_{m\neq k} \frac{H_{AB,km}}{E^{A,B}{}_m - E^{A,B}{}_k} \quad \textit{and} \quad H_{AB,km} = \left(\psi_k{}^{A,B}\middle|H_{AB}\middle|\psi_m{}^{A,B}\right) \quad \textbf{(51)}$$

The second-order approximation does not provide any new principal insight, since only quadratic terms additionally appear:

$$C_{2,k} = \sum_{A,B}\sum_{m,n\neq k} [\frac{H_{AB,km} \cdot H_{AB,kn}}{(E^{A,B}{}_m - E^{A,B}{}_k)\cdot(E^{A,B}{}_m - E^{A,B}{}_n)} - \frac{H_{AB,km} \cdot H_{AB,kn}}{(E^{A,B}{}_m - E^{A,B}{}_k)^2} + \sum_{A,B}\sum_{m\neq k} \frac{H_{AB,km}}{E^{A,B}{}_m - E^{A,B}{}_k} \cdot C_{1,k}\ ] \quad \textbf{(52)}$$

What are the implications of these results? We can verify that perturbation theory has many advantages in biochemical problems. The disadvantages of the approach (point 1) is easy to see: It is rather hopeless to compute the total systems 'drug-bioreceptor' or 'biomolecule-biomolecule', and 'bioreceptor' can be associated with a large biomolecule (DNA, RNA, protein, hormone). This approach is already hopeless, if one wishes to define the Hamiltonian of such a bioreceptor (one may think of the very complicated geometry of the double-stranded DNA including the H bonds between the base pairs interacting with chromatin). Therefore we have to restrict ourselves to theoretical means according to point 2, which also permit to use experimental properties (e.g. measurements of the ionization energy, excited states and transition probabilities inclusive intersystem crossings due to spin-orbit and spin-spin coupling). With respect to such a starting-point the methods of perturbation theory appears to be appropriate. We have already mentioned the correlation of Birks (dipole-dipole interaction between tryptophan and any carcinogen) and the role of the excited states of the PAHCs in the domain 3.1 eV – 3.5 eV (Figures 8a – 8c). This may not be a contingency, since the lowest excited states of the nucleic acids also lie in this domain: guanine (G: 3.3 eV), adenine (A:

3.35 eV), thymine (T: 3.25 eV), cytosine (C: 3.45 eV) and uracil (U: 3.17 eV).

It is known that the triplet states are unaffected by the keto-enol tautomers, and in DNA or RNA we rather observe an energy band of triplet states within the above mentioned interval than separate energy levels (triplet conduction band, Barenboim et al 1969). The first excited singlet states are more influenced by the tautomeric equilibrium induced by the H bonds between certain base pairs. Therefore these excited states lie in the large domain between 3.7 eV and 4.7 eV. It should be mentioned that tryptophan possess two excited states between 4 eV and 4.3 eV, and the lowest triplet state is in the same interval as the triplet states of the nucleic acids. This property is also true for the derivates melatonin and serotonin, but the other amino acids (inclusive phenylalanine) only possess singlet and triplet states beyond 4.7 eV and 3.6 eV, respectively. At this position it should be pointed out that by accounting for the matrix elements of ground state and excited states interactions and transition probabilities we are able to explain the chemical affinity between external molecule and bioreceptors such as DNA.

The Hamiltonian for a charged particle (e.g. electron, proton) in an external magnetic field and for spin-orbit/spin-spin couplings has the form (details are given by Hameka 1962a, 1962b, 1965, Ulmer and Hartmann 1978):

$$H_{magnetic} = \frac{1}{2\mu}\sum_{k=1}^{n} (-i\hbar\nabla_k - \frac{e}{c}\cdot\vec{A}_k)^2 \qquad (53a)$$

$$\vec{B} = \nabla\times\vec{A} \qquad (53b)$$

$$H_{I,spin} = \sum_{k=1}^{n} \underbrace{\frac{\hbar\vec{\sigma}_k}{4\mu^2c^2}\cdot(grad_k \cdot V_{\text{int}} \times \vec{p}_k)}_{(I)} + \underbrace{\frac{\hbar e}{2\mu c}\vec{\sigma}_k\cdot\vec{B}_k}_{(II)} + \underbrace{\sum_{l\neq k}^{n} T_{kl}\cdot\vec{B}_k\vec{\sigma}_k\vec{B}_l\vec{\sigma}_l}_{(III)} \qquad (54)$$

The consequences of equations (53a, 53b, 54) are tremendous, since all magnetic interactions are accounted for, and the energy levels obtained by the Hamiltonians $H_A$ and $H_B$ are additionally splitted up. In contrast to usual applications, the mass $\mu$ in equation (53a) is related to a proton mass.

With $B_z = B_0$ ($A_x = -B_0\cdot y$, $A_y = A_z = 0$) the solution function of the Schrödinger equation (53a) for free particles in a static magnetic field exhibits the form (Hameka 1965):

$$\psi_{n,\beta} = \int_{-\infty}^{\infty} f(\alpha)\exp(i(\alpha \cdot x + \beta \cdot z))\exp(-0.5 \cdot \xi^2) H_n(\xi) \qquad (55a)$$

$$E_n = \frac{e \cdot \hbar \cdot B_0}{\mu \cdot c} \cdot (n + \tfrac{1}{2});\ n = 0,1,2,....;\ \xi = (y + \frac{\hbar \alpha c}{e B_0} \cdot \sqrt{\frac{\mu \cdot \omega_0}{\hbar}}) \qquad (55b)$$

$$\omega_0 = \frac{e \cdot B_0}{\mu \cdot c} \quad (\textit{Larmor frequency}) \qquad (55c)$$

The function f(α) represents an arbitrary function; $H_n$ refers to Hermite polynomials of the degree n with the energy eigenvalue $E_n$ according to equation (55b). Please note that for electromagnetic waves equation (53a) is the basis for electric dipole transitions.

The expressions (I) – (III) have the following meanings: (I) refers to spin-orbit couplings of electrons in molecules, in cases to be explicitly mentioned (I) may also refer to protons. Expression (II) represents the spin-Pauli effect, i.e. the (small) interaction energy of the spin in an external magnetic field $B_0$). With regard to additional electromagnetic waves (II) is responsible for transitions between different spin states induced by the waves. Expression (III) refers to spin-spin coupling (electron spin, nuclear spin); it is usually very small and implies correspondingly small splittings of the energy levels.

Since all contributions (I) – (III) described by equation (54) certainly represent perturbations and splittings of discrete energy levels result from these properties, we state now the most important matrix elements (the Hamiltonian of spin-orbit coupling is denoted by $H_{SO}$):

$$D_{k,ni} = -\frac{e}{c} \cdot \sum_l \left( {}^1\psi_l \middle| H_{so} \middle| {}^3\psi_{n,i} \right) \cdot \left( {}^1\psi_k \middle| X \middle| {}^1\psi_l \right) \frac{1/}{[{}^1\varepsilon_l - {}^3\varepsilon_k]}$$
$$- \frac{e}{c} \cdot \sum_m \left( {}^3\psi_{m,i} \middle| H_{so} \middle| {}^1\psi_k \right) \cdot \left( {}^3\psi_{m,i} \middle| X \middle| {}^3\psi_{n,i} \right) \frac{1/}{[{}^1\varepsilon_m - {}^3\varepsilon_k]} \qquad (56)$$

If we look at the denominators of equation (56) we again find the similar property as valid for the electronic interactions expressed by $H_{AB}$, namely the energy difference has to be very small to record significant contributions of singlet-triplet transitions. The relevance of the term schemes 8b and 8c and their importance for the chemical reactivity are explained by these properties of the denominators. In contrast to the very fast pure singlet transitions the lifetime of triplet states has a particular meaning in long molecular chains, since charge and energy transport mechanisms are excited, which can lead to soliton transport in chains (e.g. muscles, Davydov 1979) or activate H bonds.

The second contribution (II) may represent the interaction energy of the spin magnetic moment in an external magnetic field, i.e. a given energy level E is splitted up to yield:

$$\left.\begin{aligned} E &\rightarrow E \pm \mu_{magn} \cdot B_0 \\ \mu_{magn} &= \gamma \cdot s \cdot \hbar \end{aligned}\right\} \quad \textbf{(57)}$$

The magnetic moment $\mu_{magn}$ may either refer to the proton or electron moment, $\gamma$ is the gyro-magnetic ratio. Contribution (II) also becomes relevant with regard to transitions between different spin states induced by electromagnetic waves; an example is NMR. The importance in the case of the much weaker geomagnetic field will be discussed later. The third contribution (III) refers to the coupling of spin with environmental spin systems and represents a transport mechanism with extremely low energy.

### 2.9.3. Consequences of the resonance denominators, interactions with magnetic fields and couplings to spin systems to problems of chemical affinity, H bonds and circuits

It is a general feature that all contributions (I) – (III) can be treated by well elaborated perturbation calculations, where in every order the energy differences appear in the denominators. The basic starting-point is incorporated by the Hamiltonian H and its constituents $H_A$, $H_B$, $H_{AB}$. Thus these constituents may belong to many-particle Schrödinger equations (they can be found in many textbooks of quantum mechanics, see e.g. Hameka 1965) and/or nonlinear/nonlocal Schrödinger equations with internal structure (Ulmer 1980a). The latter method also offers the possibility to calculate via charge densities and currents the properties of electric circuits in an easy manner. Furthermore, this approach incorporates in a quite natural way the density functional formalism, which has been put forward for the calculation of many-particle problems in biochemistry and molecular biology (e.g. the calculation of tunneling of protons between DNA base pairs, Pérez et al 2010). The basis skeleton of the following calculations incorporates Figures 9 and 10. The band structure (singlet and triplet states) of DNA is determined by the π electrons of the bases and the 3d-electrons of the phosphate ester (Ulmer 1979, 1980a, 1981). These 3d-electrons exhibit a large coupling range of ca. 3.5 Å, which reaches besides neighboring π electrons of the same strand also 3d-electrons of the complementary strand, inclusive π electrons of the related nucleotides. A consequence of these properties is that the upper zone of the triplet band (containing a huge number of discrete triplets) and the lower edge of the singlet excitation band overlap; a further essential property is the spin-orbit coupling of the 3d-electrons, which allows transitions from singlet to triplets and reverse. The arrows (solid lines) in Figure 9 are also valid in reverse direction, whereas the dashed arrows indicated processes with rather little probability. Owing to the splitting up of the triplet levels of 3d-electrons, we obtain an additional cascade of triplet states, which can serve as a pumping mechanism of energy, which may have its origin in ATP, GTP, etc.

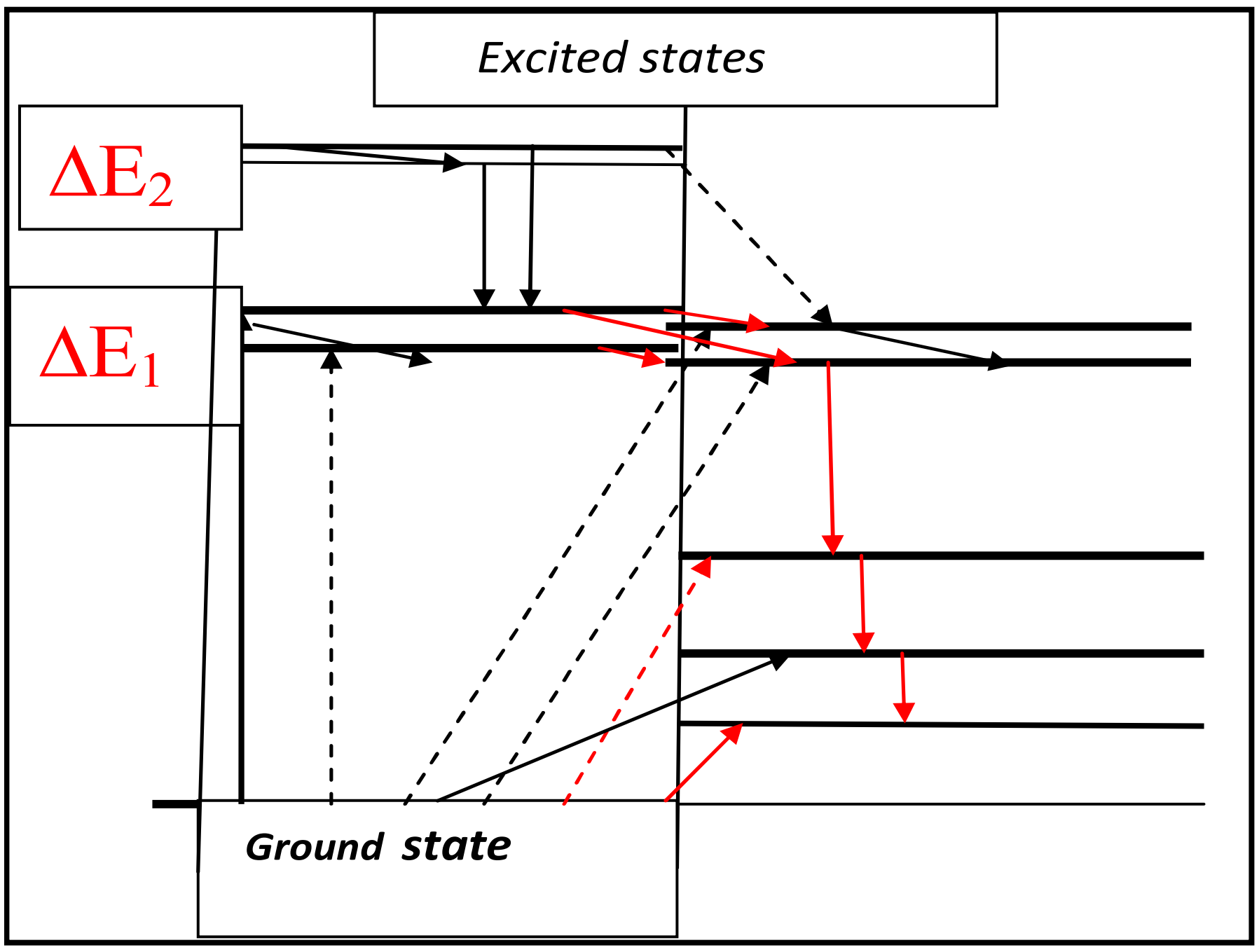

**Figure 9:** This figure includes the contents of the preceding figures 8a – 8c (singlets: left, triplets: right); the presence of one (or more) further triplet states below the double resonance open new paths for interactions, e.g. propagation of triplet states.

It is a well-known property that the H bonds between the base pairs A – T and G – C modifies the local conformation of the DNA bases according to the position of the exchange protons (keto-enol tautomers). In particular, the excited singlet states are affected by the tautomerization, but with regard to triplet states the influence to band is less noteworthy (Barenboim et al 1969). A consequence of the difference in the local charge distribution of the keto – and enol tautomer is an asymmetric potential of the H bond, which is lowered in the keto conformation. Assuming a local temperature of 300 °K the proton tunneling velocity from the right-hand side (enol tautomer) of Figure 11 to the left-hand (keto tautomer) is fast, whereas for the reverse process the corresponding velocity is extremely small (factor $10^3$). Since the charged tunneling protons incorporate spin ½ and a current between the corresponding base pairs, we can treat the energetic processes as perturbations. The influence of the geomagnetic field (order $0.5 \cdot 10^{-5}$ Tesla) is rather difficult but very promising due to the helical structure of DNA:

1. If the direction of tunnel current between two base pairs is perpendicular to the direction of the geomagnetic field, then we have a maximum effect of the superimposed rotational motion induced by the Lorentz force (highest Larmor frequency).

2. If the direction of tunnel current between two base pairs is (approximately) parallel to the direction of the geomagnetic field, then we have a minimum effect of the superimposed screw induced by the Lorentz force (lowest Larmor frequency). Geomagnetic and solar magnetic fields have now about the same order of strength.

3. There are further configurations of H bonds between base pairs, where the projection of the magnetic induction $B_0$ lies between the extreme cases 1 and 2.

The rotational motion of the protons induces additional spin-orbit couplings and spin-spin couplings with neighboring H bonds and 3d-electrons. The extremely small energy differences in the related dominators of the perturbation expansion of the wave functions yield sensitive resonances by external electromagnetic waves (with origins from the earth or the sun). The possible energy levels (singlets, triplets) of DNA suffer various further splitting, which can be excited by appropriate resonance conditions. In the language of electromagnetic circuits spin-orbit and spin-spin couplings can be interpreted as mutual inductance M (cases (I) and (III) of equation (54)). The interaction energy of the proton spin in the geomagnetic field is represented by the contribution (II) of this equation (spin-Pauli effect). The Lorentz force (interaction of the charged proton with the geomagnetic field) is expressed by the Hamiltonian (53a). This motion is quantized and leads to discrete energy levels expressed by the Larmor frequency (ground frequency). Excitations of the ground state can only occur by adsorption of electromagnetic waves with very low energy and proper eigenfrequency.

The asymmetry of the potential between DNA base pairs according to the dashed line of Figure 11 exhibits an additional interesting consequence. Due to the long probability of abidance in the keto conformation compared to the abidance in the enol conformation the helical DNA assumes an additional stability with regard to mutagenic influences. If a potential according to model 1 would be valid, then this property would not exist and the motion of a proton between the two minima due to quantum mechanical tunneling would be identical.

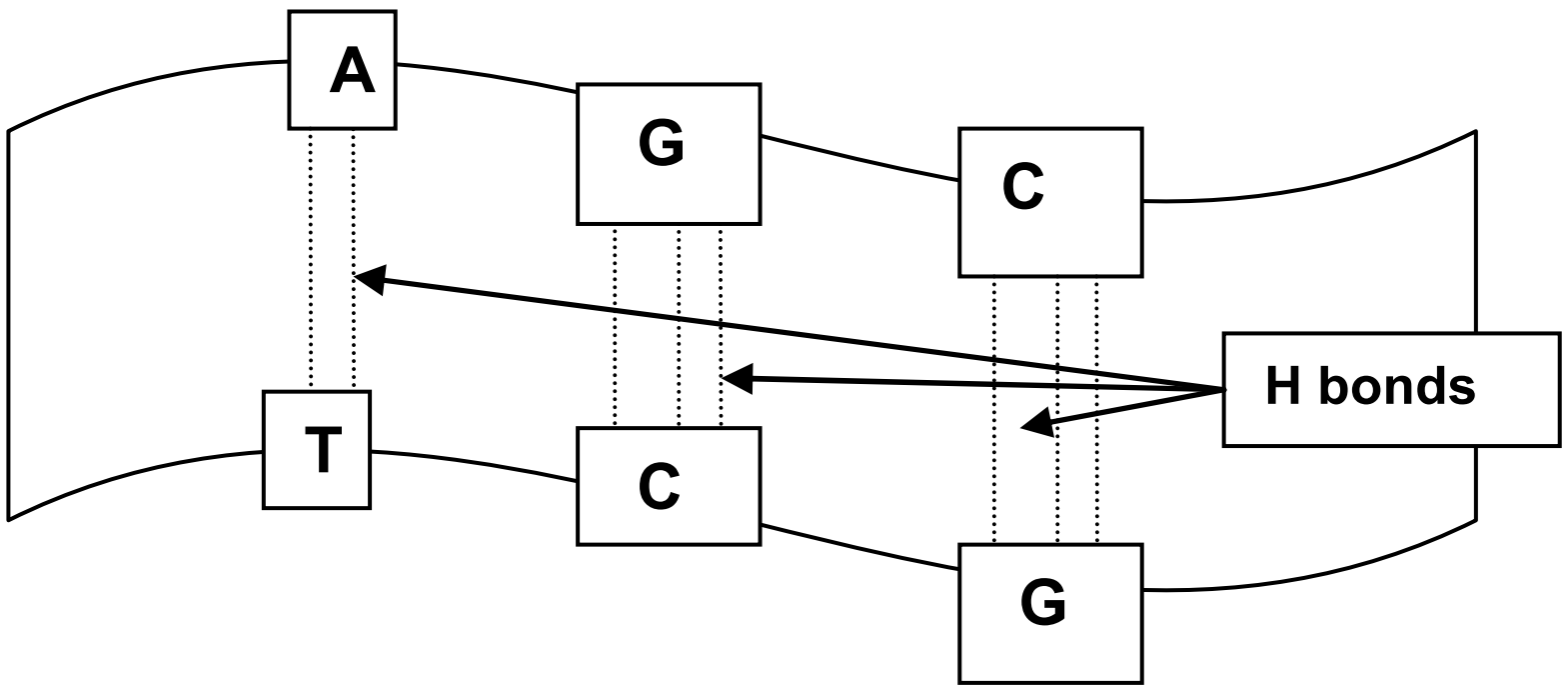

**Figure 10:** Section of the double-stranded DNA helix and the H bonds (protons) between the base pairs A – T and G – C.

The additional circular motion of protons between base pairs seems to play also an important role in DNA replication and transcription, since it represents a key for the rotation specific parts of the DNA strands. Some consequences of calculations referring to this section will be discussed in section 3.

Figure 11 is obtained by a nonlinear/nonlocal field with internal structure and spin (Ulmer 1980a):

$$\left.\begin{aligned}
& i\hbar\cdot\frac{\partial}{\partial t}\cdot\psi = \frac{1}{2\mu}[\vec{\sigma}[-i\hbar\nabla - \frac{e}{c}\cdot\vec{A}]]^2\psi - H_{so}\cdot\psi + \lambda\cdot\vec{\sigma}\cdot\int(\psi^+(\vec{x}\,')\vec{\sigma}\psi(\vec{x}\,')K(\vec{x}-\vec{x}\,')d^3x'\cdot\psi(\vec{x}) \\
& K(\vec{x}-\vec{x}\,') = \frac{1}{\sqrt{\pi}^3\cdot\varepsilon_1\cdot\varepsilon_2\cdot\varepsilon_3}\cdot\exp(-(x-x')^2/\varepsilon_1{}^2 + (y-y')^2/\varepsilon_2{}^2 + (z-z')^2/\varepsilon_3{}^2)\cdot \\
& \qquad \cdot\sum_{l=0}^{\infty}\sum_{m=0}^{\infty}\sum_{n=0}^{\infty} P_{l,m,n}\cdot H_l((x-x')/\varepsilon_1)\cdot H_m((y-y')/\varepsilon_2)\cdot H_n((z-z')/\varepsilon_3)
\end{aligned}\right\}\quad (57)$$

In lowest order, the solution functions can be approximated by the product wave functions

$$\psi = \phi_{electrical\ \ part} \cdot \Phi_{magnetic} \quad (57a)$$

Inclusion of higher order contributions of magnetic perturbations to the leading electrical part of the Hamiltonian $\phi_{electrical\ \ part}$ again implies resonance dominators as already pointed out for $H_{AB}$.

The necessary integration procedure has already been worked out (Ulmer 1980a); magnetic interactions by external fields and interactions with spin are treated as small perturbations. The realistic potential for tunneling protons between DNA base pairs is easy to handle by the generalized Gaussian convolution kernel K, which contains multipole expansions in arbitrary order accounted for in terms of two-point Hermite polynomials $H_n$.

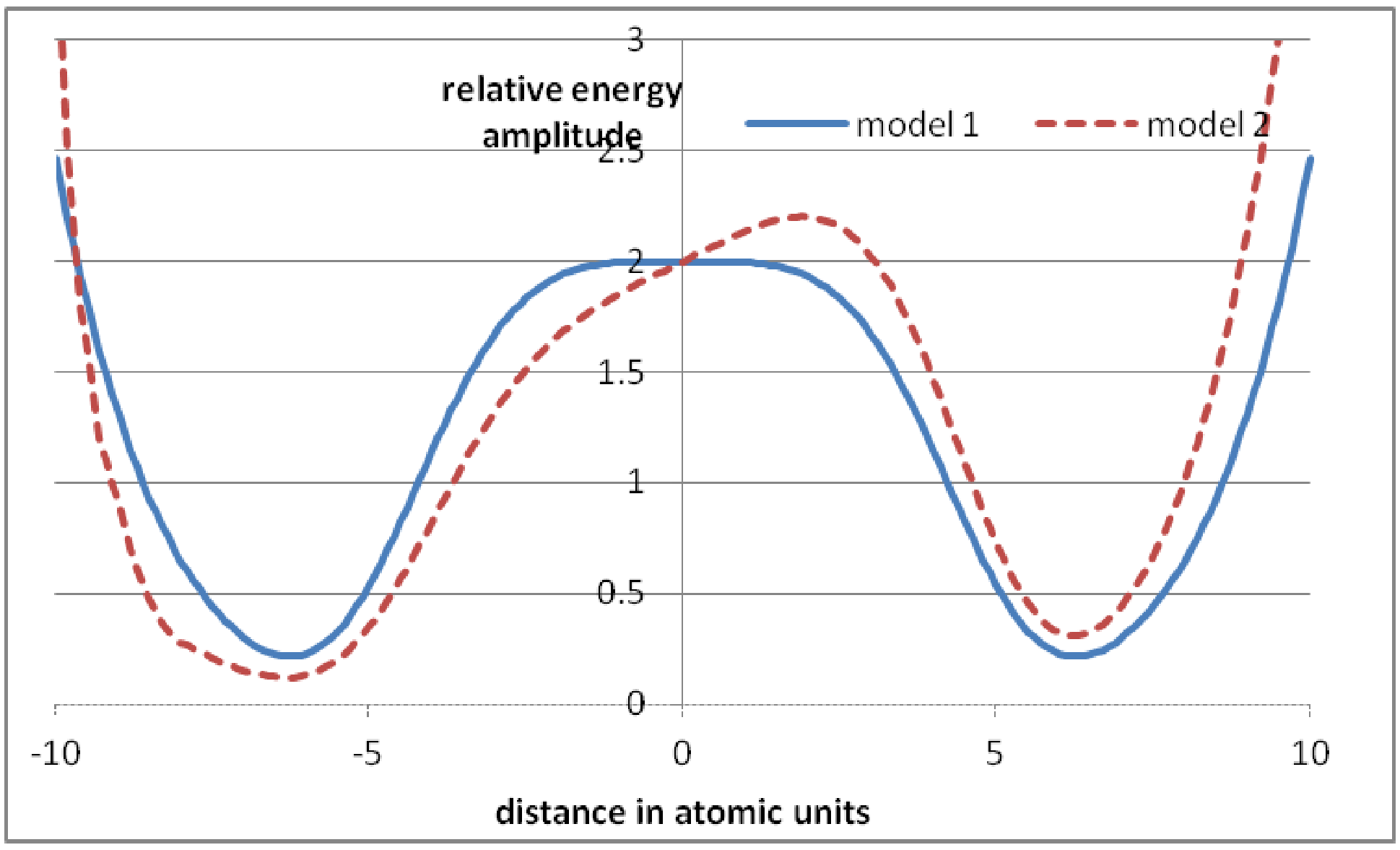

**Figure 11:** Double minimum potential between the base pairs A – T. Left minimum: Keto-tautomer of A (dashed line, model 2). Model 1 represents the symmetric approximation is valid for H bonds in water.

## 2.10. Connection to diffusion theory of charged particles in magnetic fields

In this section, we discuss the problem of the interaction of charged particles from a different viewpoint, namely the reaction-diffusion theory, which is also valuable as a completion of the preceding sections.

### 2.10.1. Fick's law of diffusion and the Kolmogorov forward equation

Diffusion (i.e. Brownian motion) has the general property that concentrations of particles/molecules tend to distribute every available space, if there are no constraints. Fick's law of diffusion plays a key role in a lot of physiological processes, since the restriction to pure biochemical kinetics requires a spatial equilibrium of concentrations, which does not exist. Fick's law is well-known, and it shows the same structure as other transport equations, i.e. transport of heat:

$$-\frac{\partial}{\partial t}\rho + D\,\Delta\,\rho = 0 \quad (58)$$

D is the diffusion coefficient, Δ the Laplace operator and ρ the particle density. This equation results from the property that a concentration gradient implies a current, and a balance equation (conservation of the overall particle number) must be satisfied in the absence of chemical potentials:

$$\left.\begin{array}{l} \vec{j} = -D\nabla\rho \\ \frac{\partial}{\partial t}\rho + div\,\vec{j} = 0 \\ div\,\vec{j} = (\nabla\cdot\vec{v})\rho \end{array}\right\} (59)$$

If the particles are ions with electric charge q, then in the presence of magnetic fields, described by the vector potential **A**, an additional momentum due to a magnetic field (Lorentz force) has to be accounted for:

$$\left.\begin{array}{l} \vec{p} = -\frac{q}{c}\vec{A} \\ \vec{v} = \vec{p}\,/\,M = -\frac{q}{c\,M}\vec{A} \\ div\;\vec{A} = \nabla\cdot\vec{A} = 0 \end{array}\right\} (60)$$

M: particle mass, which may also be an effective mass, c: velocity of light, p: momentum. Inserting (60) in equation (59) we obtain following equations:

$$\left.\begin{array}{l} \vec{j} = -(D \cdot \nabla - \frac{q}{MC}\vec{A})\rho \\ \frac{\partial}{\partial t}\rho + div\ \vec{j} = 0 \\ -\frac{\partial}{\partial t}\rho + D \cdot \Delta\rho - \frac{q}{MC}(\vec{A} \cdot \nabla)\rho = 0 \end{array}\right\} (61)$$

From equation (61) follows the Kolmogorov forward equation:

$$-\frac{\partial}{\partial t}\rho + D \cdot \Delta\rho - div\,(v \cdot \rho) = 0 \quad (62)$$

The term div(**v**·ρ) results from the action of a magnetic field and is regarded as a directional force, whereas pure diffusion has no preference direction. In above equations we have only modified the diffusion current by the magnetic interaction, but not yet the balance equation due to the presence of a potential. In order to reach full gauge invariance we have to write:

$$\left.\begin{array}{l} \frac{\partial}{\partial t}\rho + (\nabla - \frac{a}{McD}\vec{A})(D \cdot \nabla - \frac{q}{MC}\vec{A}) \cdot \rho = 0 \\ -\frac{\partial}{\partial t}\rho + D \cdot \Delta\rho - \frac{2q}{Mc}(\vec{A} \cdot \nabla)\rho + (q^2\,\vec{A}^2 /[D \cdot M^2 \cdot c^2]) \cdot \rho = 0 \end{array}\right\} (63)$$

This equation shows a particular importance, since it can be transformed to a Schrödinger equation with magnetic field (53a). The Kolmogorov forward equation (60) is a special case of equation (63).

## 2.10.2. External magnetic field with $B_0$ in z-direction

At first, we consider the case $B_z = B_0$ ($B_x = B_y = 0$, **B** = curl**A**), we choose **A** as follows:

$$\left.\begin{array}{l} A_x = -B_0 \cdot y;\ B_z = \partial A_y/\partial x - \partial A_x/\partial y = B_0 \\ \mathbf{A}^2 = B_0^2 \cdot y^2 \end{array}\right\} (64)$$

Inserting equation (64) in equation (63) and using the substitution $\omega_0 = \frac{q \cdot B_0}{M \cdot c}$ we obtain:

$$-\frac{\partial}{\partial t} \cdot \rho + D \cdot \Delta\rho + 2 \cdot \omega_0 \cdot y \cdot \frac{\partial}{\partial x} \cdot \rho + \frac{\omega_0^2}{D} \cdot Y^2 \cdot \rho = 0 \quad (65)$$

Equation (66) is solved by the following procedure:

$$\rho(x,y,z,t) = \varphi(y) \cdot \exp(i \cdot \alpha \cdot x) \cdot \exp(i \cdot \gamma \cdot z) \cdot \exp(-D \cdot t \cdot (\alpha^2 + \gamma^2)) \quad (66)$$

The contribution from the z-direction leads to pure diffusion, whereas the x-direction is also connected to the y-direction due to the term $2 \cdot \omega_0 \cdot y \cdot \frac{\partial}{\partial x} \cdot \rho$ in equation (57). Therefore φ(y) is determined by the equation:

$$\frac{\partial^2}{\partial y^2} \cdot \varphi + \frac{2 \cdot i \cdot \alpha \cdot \omega_0}{D} \cdot y \cdot \varphi + \frac{\omega_0^{\,2}}{D^2} \cdot y^2 \cdot \varphi = 0 \quad (67)$$

The following substitutions are performed:

$$\left.\begin{array}{l} y_s = y + \eta; \;\; \eta = i \cdot \alpha \cdot D / \omega_0 \\ \xi = y_s \cdot \sqrt{\frac{i \cdot \omega_0}{D}} \end{array}\right\} \quad (68)$$

This provides the dimensionless basic equation

$$\partial^2 \varphi / \partial \xi^2 + (\delta - \xi^2) \cdot \varphi = 0 \quad (69)$$

The solution functions are a complex Gaussian multiplied with Hermite polynomials in the complex space:

$$\left.\begin{array}{l} \varphi_n = H_n(\xi) \cdot \exp(-\xi^2 / 2) \\ \delta = 2 \cdot n + 1 \;\; (n = 0,1,....); \;\; \alpha(n)^2 = i \cdot (2n+1) \cdot \omega_0 / D \end{array}\right\} \quad (70)$$

With regard to relations (69) and (70) some comments are justified. The solution function $\varphi_n$ formally agrees with that of quantum mechanics (oscillator in a magnetic field, $\omega_0$: Larmor frequency), but $\xi$ is not real (see (68)). According to trigonometric theorems in the complex space it is also possible to represent $\varphi_n$ by combinations of sine – and cosine functions. Therefore the solution function of equation (55) is completed by the procedure:

$$\rho = \Phi(z,t) \cdot \varphi_n(y) \cdot \exp(i \cdot \alpha(n) \cdot x) \cdot \exp(-D \cdot t \cdot \alpha(n)^2) \quad (71)$$

The solution function φ(y) satisfying equation (67) can be directly obtained by a sum of a sine - and cosine function, if n = 0 i.e. $exp(-\xi^2 / 2)$ is transformed to partially real functions.

$$\varphi(y) = A \cdot \sin(P_0 \cdot y^2 + P_1 \cdot y) + B \cdot \cos(P_0 \cdot y^2 + P_1 \cdot y) \quad (72)$$

In the case of n > 0 we have to multiply A and B with polynomials of y and to determine the coefficients via equation (67). Inserting equation (72) into equation (67) we obtain:

$$P_0 = \pm \frac{\omega_0}{2 \cdot D}; \quad P_1 = \pm i \cdot \alpha \quad (73)$$

So the sine and cosine still contain a complex argument. The magnitude of A and N depends of boundary/initial conditions, but in complicated systems like molecular biology they are hardly to determine. There are basically two restrictions, given by equations (67, 72 – 73):

$$\left.\begin{array}{l} 1.\ A = -B \\ 2.\ B = A \cdot \dfrac{D \cdot \alpha^2 + \omega_0}{D \cdot \alpha^2 - \omega_0} \end{array}\right\} \quad (74)$$

For the general case with polynomials of y and n ≥ 0, formula (74) has to be modified:

$$3.\ B = A \cdot \frac{D \cdot \alpha^2 + \omega_0 \cdot (2 \cdot n + 1)}{D \cdot \alpha^2 - \omega_0 \cdot (2 \cdot n + 1)} \quad (75)$$

The denominator of equations (74 – 75) seems to be noteworthy, since it may vanish or, at least, very huge. This behavior corresponds to a resonance condition. We shall return to this aspect after an analysis of the geomagnetic and solar magnetic field.

According to previous results $\Phi(z,t)$ obtained by carrying out the integral

$$\Phi = \int_{-\infty}^{\infty} \exp(i \cdot \gamma \cdot z) \cdot \exp(-D \cdot t \cdot \gamma^2) f(\gamma) d\gamma \quad (76)$$

If $f(\gamma) = 1$, then $\Phi(z,t)$ is given by the well-known Gaussian, i.e.:

$$\Phi(z,t) = \frac{1}{\sqrt{4 \cdot \pi \cdot D \cdot t}} \exp(-z^2 / 4 \cdot D \cdot t) \quad (77)$$

If $f(\gamma)$ is given by a polynomial expansion, then we obtain a Gaussian multiplied with certain polynomials $P_l$ (Ulmer 1983):

$$\Phi_l(z,t) = P_l(z,t) \cdot \frac{1}{\sqrt{4 \cdot \pi \cdot D \cdot t}} \exp(-z^2 / 4 \cdot D \cdot t) \quad (78)$$

The calculation of $P_l$ is developed in the cited reference. On the other side, if the space available for diffusion has constraints and boundaries, then $\Phi(z,t)$ is determined by some error functions erf. The spatial boundaries may be given by $z_1 = 1/\gamma_1$ and $z_2 = 1/\gamma_2$, and we now obtain:

$$\left.\begin{array}{l}\Phi=U(t)\cdot\exp(-z^2/4\cdot D\cdot t)\cdot[erf(s_1)-erf(s_2)]\\ U(t)=\dfrac{z_1+z_2}{8\sqrt{2\cdot D\cdot t}}\\ s_1=(1/z_1+i\cdot z/2\cdot D\cdot t)/\sqrt{D\cdot t};\;\; s_2=(-1/z_2+i\cdot z/2\cdot D\cdot t)/\sqrt{D\cdot t}\end{array}\right\}\quad(79)$$

It can be seen that due to the complex values of $s_1$ and $s_2$ the diffusion of the charged particles adopts an oscillatory behavior resulting from the constraint boundaries. In every case, the behavior in x-direction is determined by the magnetic field, since α(n) is already fixed by equation (70). This fact implies that both **exp(i·α·x)** and **exp(-D·α(n)$^2$·t)** exhibit complex values, i.e. we can obtain either damped oscillations or enhanced oscillations in dependence of the actual concentrations and connections with possible pools. We point out those superpositions with different values of n are also allowed.

It is a straightforward task to extend the basic equation (63) by a reaction potential at the right-hand side, which is given by either

1. $\lambda_1\cdot\rho$ (decay of the concentration ρ due to the coupling to a further component).

2. $\lambda_1\cdot(1-\rho)$ (formation of the concentration ρ due to the decay-coupling of a further component).

The general solution (79) has to be generalized (decay/formation) reaction:

$$\left.\begin{array}{l}\rho=\Phi(z,t)\cdot\varphi_n(y)\cdot\exp(i\cdot\alpha(n)\cdot x)\cdot\exp(-D\cdot t\cdot\alpha(n)^2)\cdot\exp(-\lambda_1\cdot t)\\ (decay)\\ \rho=\Phi(z,t)\cdot\varphi_n(y)\cdot\exp(i\cdot\alpha(n)\cdot x)\cdot\exp(-D\cdot t\cdot\alpha(n)^2)\cdot(1-\exp(-\lambda_1\cdot t))\\ (formation)\end{array}\right\}\quad(80)$$

This result is of interest with regard to the x-axis contribution, i.e., $\exp(i\cdot\alpha(n)\cdot x)\cdot\exp(-D\cdot t\cdot\alpha(n)^2)$, since both exponential functions exhibit oscillatory behavior, and either enhancement or prohibition by damping are possible consequences. This is also an example of the directional properties of magnetic fields in chemical reactions.

We shall now stress our interest to the geomagnetic and solar magnetic field. Since the geomagnetic field depends on the localization on the earth surface, a separation of both components makes only sense, if we consider that projection of the solar magnetic field, which is perpendicular to the geomagnetic direction (solar magnetic field: $B_z = B_0$; geomagnetic field:

$B_x = B_1$). As the geomagnetic field strength is about $10^4$ ·solar magnetic field strength, the projection of the solar field in direction to the geomagnetic field is completely negligible. Since the blood flow also carries ions, the magnetic field strength of this current induces a magnetic field of the order of the solar magnetic field, and the projection of that component, which has the same orientation as the solar magnetic field, may slightly enhance the solar magnetic field strength. With regard to the vectors **A** (vector potential) and **B** (magnetic induction) we use the following definitions:

$$A = \begin{pmatrix} 0 \\ B_0 \cdot x \\ B_1 \cdot y \end{pmatrix} \text{ (81 ) and } B = \begin{pmatrix} B_1 \\ 0 \\ B_0 \end{pmatrix} \text{ (82)}$$

Now equation (63) assumes the shape:

$$\left.\begin{array}{l} -\frac{\partial}{\partial t} \cdot \rho + D \cdot \Delta \rho - 2 \cdot \omega_0 \cdot x \cdot \frac{\partial}{\partial y} \cdot \rho - 2 \cdot \omega_1 \cdot y \cdot \frac{\partial}{\partial z} \cdot \rho + \frac{\omega_0{}^2 \cdot x^2}{D} \cdot \rho + \frac{\omega_1{}^2 \cdot y^2}{D} \cdot \rho = 0 \\ \omega_0 = \frac{q \cdot B_0}{M \cdot c}; \ \omega_1 = \frac{q \cdot B_1}{M \cdot c} \end{array}\right\} \text{(83)}$$

We modify formula (66) to solve above equation (83) nearly exact:

$$\rho = \Phi(x, y) \cdot \exp(-i \cdot \gamma \cdot z) \cdot \exp(-D \cdot t \cdot \gamma^2) \text{ (84)}$$

Thus we obtain:

$$D \cdot \left(\frac{\partial^2}{\partial x^2} + \frac{\partial^2}{\partial y^2}\right) \cdot \Phi - 2 \cdot \omega_0 \cdot x \cdot \frac{\partial}{\partial y} \cdot \Phi - 2 \cdot i \cdot \gamma \cdot \omega_1 \cdot y \cdot \Phi + \left(\frac{\omega_0{}^2 \cdot x^2}{D} + \frac{\omega_1{}^2 \cdot y^2}{D}\right) \cdot \Phi = 0 \text{(85)}$$

In order to make use of equations (68 – 70) with respect to the y-coordinate, we have, at first, to neglect the term $2 \cdot \omega_0 \cdot x \cdot \frac{\partial}{\partial y} \cdot \Phi$, since $\omega_1 >> \omega_0$ ($B_1 >> B_0$). In a second step we shall account for this term, which incorporates a correlation between both components. Therefore we obtain:

$$\left.\begin{array}{l} y_s = y - \eta; \ \eta = i \cdot \gamma \cdot D / \omega_1 \\ \xi = y_s \cdot \sqrt{\frac{i \cdot \omega_1}{D}} \end{array}\right\} \text{(86)}$$

This yields formally the harmonic oscillator equation of quantum mechanics:

$$\left.\begin{array}{l}\partial^2\varphi_2/\partial\xi^2+(\delta-\xi^2)\cdot\varphi=0\\ \varphi_{2,n}=H_n(\xi)\cdot\exp(-\xi^2/2)\\ \delta=2\cdot n+1\ \ (n=0,1,....);\ \gamma(n)^2=i\cdot(2n+1)\cdot\omega_1/D\end{array}\right\}\ (87)$$

Finally we have to consider $\varphi_1(x)\cdot f(x, y)$; f(x, y) results from the product **x·d/dy** of equation (85). In a first order, we take account of the 'mixed product' by the function

$$f=\exp(\tfrac{1}{2}\cdot i\cdot x\cdot y\cdot\omega_0/D)(88)$$

Then we have to solve the equation

$$\frac{d^2}{dx^2}\cdot\varphi_1+\frac{\omega_0{}^2\cdot x^2}{D^2}\cdot(1-i\cdot\exp(\tfrac{1}{2}\cdot i\cdot x\cdot y\cdot\omega_0/D))\cdot\varphi_1=0\quad(89)$$

This formula can be iteratively solved to yield:

$$\left.\begin{array}{l}\varphi_{1,0}=A\cdot\sin(p\cdot x^2)+B\cdot\cos(p\cdot x^2)\\ p=\pm\tfrac{1}{2}\cdot\omega_0/D\end{array}\right\}\ (90)$$

In the next step we have to account for the 'mixed product exponential function', which in physics is referred to as perturbation expansion:

$$\varphi_{1,1}=\varphi_{1,0}\cdot[1+h_1(x,y)]\quad(91)$$

Repeating this procedure we obtain:

$$\left.\begin{array}{l}\varphi_1(x,y)=\varphi_{1,0}\cdot[1+h(x,y)]\\ h(x,y)=\sum_{m=1}^{\infty}a_m\cdot(\frac{1}{\sqrt{D\cdot\omega_0}}\cdot y\cdot\omega_0)^m\cdot\cos^m(\tfrac{1}{2}\cdot\omega_0\cdot x\cdot y/D)\end{array}\right\}(92)$$

The total solution is given by the products

$$\rho=\exp(i\cdot\gamma(n)\cdot z)\cdot\exp(-D\cdot t\cdot\gamma(n)^2)\cdot\varphi_{2,n}(y)\cdot\varphi_1(x,\ y)\quad(93)$$

There is no more any degree of freedom for pure diffusion, since even in z-direction the solutions depend on those of the y-direction. Furthermore, both x – and y – direction are modulated by h(x, y). It is evident that in x – direction, where the very slow $\omega_0$ is dominant due to $B_0$, the stochastic motion of ions is superimposed by corresponding slow oscillations. On the other hand, the fast oscillations in y –direction are modulated by the frequency $\omega_0$ and powers of $\omega_0$.

Since we cannot switch off magnetic fields in cells and all biomolecules are more or less charged besides the metallic ions, there are various implications of this fact: We particular think of the organization of biomolecules and morphology, where the directional forces may have played a significant role in the evolution.

## 3.0. Some Applications

In this section we consider problems of energy/charge transfer processes of DNA and the hydrolytic decomposition of ATP.

## 3.1. Properties of DNA

Figure 10 clearly shows that the H-bonds between DNA base pairs mediated by the protons represent a current. Since the current produced by each proton induces a magnetic field there exists a magnetic coupling. The strength of this coupling depends on phase properties, e.g. whether neighboring protons simultaneously run parallel or opposite between the base pairs. On the other side, the charge distributions in the DNA-basis coupled via deoxyribophosphates have to be regarded as coupled capacitances. The proton current between base pairs have been calculated with the help of equation (57). This current is preferably affected by the geomagnetic field (Figures 12a and 12b). Since the DNA exists as a double helix, it depends on the local orientation of the H bond related to the external field, whether $B_{geomagnetic}$ is fully or in a limit situation, very weak present due to the Lorentz force. Therefore the proton current cannot only occur along the smallest distance (straight-line) of the base pairs, but the corresponding motion represents an additional screw. The rotational frequency is characterized by the Larmor frequency. In the limit case of a very small projection angle a further crew originated by the extremely small solar magnetic field may also influence the proton motion. Therefore the overall motion leads to further couplings along the H bonds of the double helix: spin-orbit coupling according to the rotational motion and spin-spin coupling between the moving protons and between protons and 3d-electrons of the phosphates of the nucleotides. Since the DNA bases of the strands mutually change configurations interactions (mixture of singlets and triplets with respect to keto-enol tautomers), the whole system of both strands represents resonators with dielectric coupling.

With regard to the magnetic interactions along the double helix we have to account for rather weak couplings (spin-orbit and spin-spin). However, due to the H bonds mediated by tunnel currents between base pairs the wavelike motions (electric couplings between the bases of the related DNA strands and magnetic interactions) are not arbitrary; they coincide in the phases.

The properties of Figures 12a and 12b in connection with Figure 11 are most important for biorhythms. Thus the potential according to Figure 11 is with slight modifications valid for all H bonds of DNA. If this potential would have the form of a flat minimum between the related base pairs, the protons would oscillate within this minimum according to the local temperature, any H bond would be possible between the base pairs and a keto-enol tautomerization would not exist. This means that the whole double-stranded helix could not exist in the real form. Since the interaction mediated by H bonds is connected with quantum mechanical tunneling, the frequencies of the protons travelling between base pairs are significantly lowered compared to thermal oscillations in a potential minimum.

***Figure 12a***

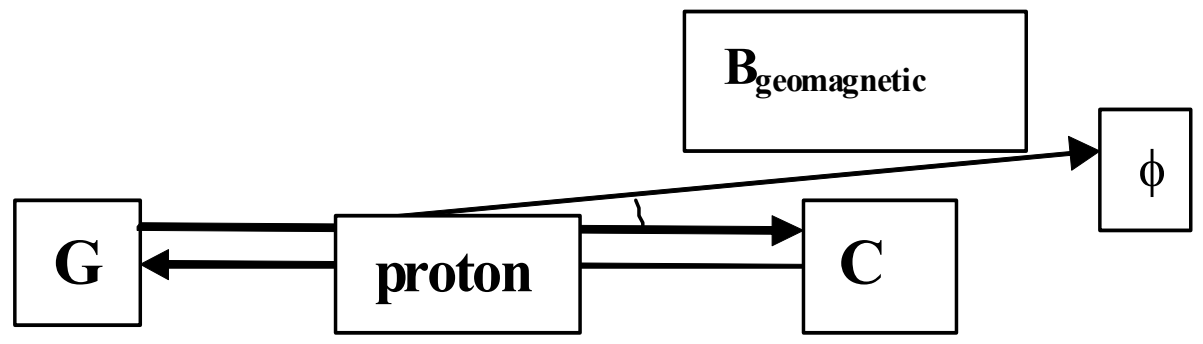

***Figure 12b***

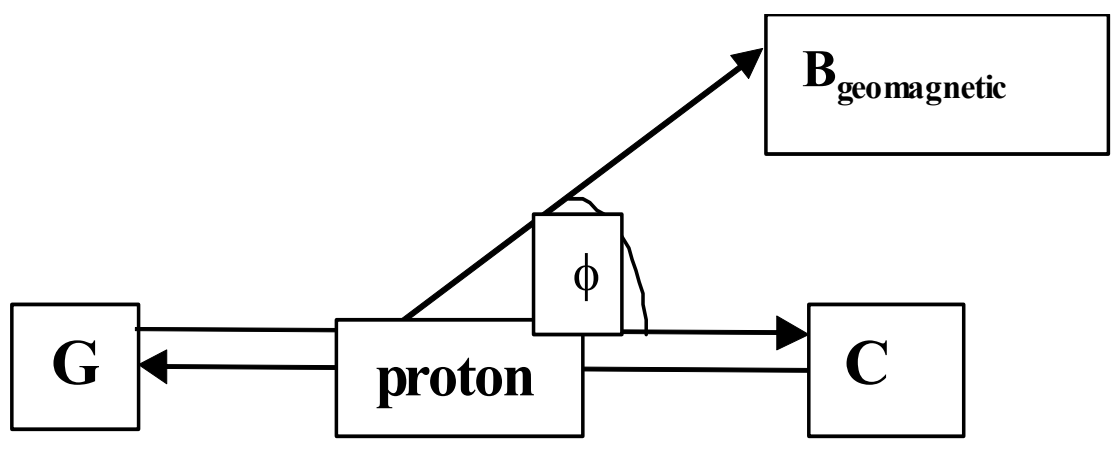

**Figures 12a and 12b:** Directions of the geomagnetic field related to an H bond between the base pairs G and C. In Figure 12a $\phi$ is extremely small (H bond and $\mathbf{B}_{geomagnetic}$ are nearly parallel; in Figure 12b the projection of $\mathbf{B}_{geomagnetic}$ perpendicular to the axis between G and C is significantly higher.

At first, let us consider the H bond between A and T, G and C in connection with Figure 11.

Thus the protons oscillate comparably fast within the corresponding potential minimum before successful tunneling can happen. This gives rise for an oscillating current expressed by inductivance $L$, the duration of halt probability at the bases is connected to the capacitance $C$. In first order, we can represent the 3 H bonds (G…C) by 3 identical oscillators (Figure 3 or Figure 6a, since they are not positioned in a coplanar way) and the 2 H bonds (A…T) by Figure 2 or Figure 4. Table 1 presents the results with/without magnetic couplings, which have been determined via calculations of the charges at the base pairs and currents between them.

**Table 1:** Resonance frequencies of the H bonds of DNA with/without coupling between the corresponding base pairs A – T and G – C.

| A – T ($2\pi/\tau$ in $sec^{-1}$) | | | G – C ($2\pi/\tau$ in $sec^{-1}$) | | | |
|---|---|---|---|---|---|---|
| $\omega_0$ | $\omega_{10}$ | $\omega_{20}$ | $\omega_0$ | $\omega_{10}$ | $\omega_{20}$ | $\omega_{30}$ |
| 6 .15·E-3 | 2.12·E-2 | 8.72·E-3 | 6.04·E-3 | 4.09·E-2 | 1.22·E-2 | 9.82·E-3 |

The next step accounts for the magnetic interaction of protons of the H bonds in the geomagnetic field (Figures 12a, b) leading to Larmor frequencies. Due to the double helix we can obtain all possibilities, i.e. a maximal effect, if proton motion and magnetic field are perpendicular, and a minimum influence, if both are parallel. In the latter case, the solar magnetic field can act with a very slow frequency. The time of resonances vary between 9 sec and 2 hours (if only the solar magnetic field is present under the particular configuration of the double helix). Since this motion exhibits always the form of circles around the axis between the base pairs, the geomagnetic field is permanently changing, which finally leads to mutual couplings between the protons travelling through and back and to spin-orbit/spin-spin couplings. A consequence of all these mutual couplings is that we obtain besides the fast oscillations ‘beat frequencies’ of the order:

$$\tau_0 \approx \mathbf{0.5}\,\boldsymbol{day};\ \tau_1 \approx \mathbf{0.95}\,\boldsymbol{day};\ \tau_2 \approx \mathbf{3.4}\,\boldsymbol{days};\ \tau_3 \approx \mathbf{6.7}\,\boldsymbol{days}\,.$$

In the continuum limit, which provides again standing waves, we have also to account for these coupling. According to a previous section the related wave equation with asymmetric/nonlocal magnetic coupling reads:

$$\left.\begin{array}{l}\partial^2/\partial t^2[1+l_c^2\cdot\partial^2/\partial x^2+Terms\;]\cdot Q-v^2\partial^2 Q/\partial x^2=0\\ l_c^2=\Delta x^2\cdot M/(L+2M)\\ Terms=a_3\cdot\partial^3/\partial x^3+a_4\cdot\partial^4/\partial x^4\end{array}\right\}\quad(94)$$

The asymmetric/nonlocal contributions results from weak couplings of longer range. Thus the coefficients $a_3$ and $a_4$ are given by:

$$a_3=\Delta x^3\cdot M/(L+3M);\; a_4=\Delta x^4\cdot M/(L+4M)\quad(94a)$$

Contributions of higher order can be neglected, but they indicate the complex situation of DNA. In similar fashion, the DNA bases along the strands are also connected in a nonlocal way, mainly due to the coupling via 3d electrons of the phosphate groups. Equation (36) now becomes:

$$\left.\begin{array}{l}(\partial^2/\partial t^2-v^2\partial^2/\partial x^2+b_3\partial^3/\partial x^3+b_4\partial^4/\partial x^4)\cdot Q\\ b_3=-\Delta x\cdot v^2;\quad b_4=-\Delta x^2\cdot v^2\end{array}\right\}\quad(95)$$

The standing wave solutions of equations (45 – 47) cannot hold in equations (94 – 95) due to the correction terms. Thus we obtain terms up to order 4, which now resemble the solutions of resonators consisting of 4 coupling terms.

Due to spin-spin coupling we have also to add the existence of spin waves, of which the spin orientation is given by the direction of the geomagnetic field (so far other magnetic of stronger order are absent). If we replace in equation (47) charge Q by spin S, an equivalent equation holds (see Hameka 1965). In a first order equation (94) provides the already stated solutions:

$$\left.\begin{array}{l}Q=c_0\cdot Q_0\cdot\exp(i\cdot\omega_n t)\cdot\sin(k\cdot x)\\ k\cdot a=m\cdot\pi\Rightarrow k=m\cdot\pi/a\quad(m=1,2,....)\end{array}\right\}\quad(96)$$

The difference to equation (45) is only the weight factor $c_0$, the eigenvalues are still unchanged:

$$\left.\begin{array}{l}n^2=m^2\pi^2 l_c^2/a^2\quad(m=1,2,....)\\ \omega_m^2=\omega_0^2/[(m^2\pi^2 l_c^2/a^2+1)\cdot(1+2\omega_0^2\cdot M\cdot C)]\end{array}\right\}\quad(96a)$$

In further orders we account for the couplings. By that, we modify equation (96) by adding the terms

(note that $c_0 + c_1 + c_2 = 1$):

$$\left.\begin{aligned}
&Q = c_0 \cdot Q_0 \cdot \exp(i \cdot \omega_n t) \cdot \sin(k \cdot x) \\
&+ c_1 \cdot Q_0 \cdot \exp(i \cdot \omega'_n t) \cdot (\sin(k' \cdot x) + \cos(k' \cdot x)) + c_2 \cdot Q_0 \cdot \exp(i \cdot \omega''_n t) \cdot \sin(k'' \cdot x) \\
&k \cdot a = m \cdot \pi \Rightarrow k = m \cdot \pi / a;\ k' = m \cdot \pi / 2a;\ k'' = m \cdot \pi / 3a \quad (m = 1, 2, ....) \\
&\omega_m{}^2 = \omega_0{}^2 / [(m^2 \pi^2 l_c{}^2 / a^2 + 1) \cdot (1 + 2\omega_0{}^2 \cdot M \cdot C) \\
&\omega'_m{}^2 = \omega_0{}^2 / [(m^2 \pi^2 l_c{}^2 / 4a^2 + 1) \cdot (1 + 2\omega_0{}^2 \cdot M_1 \cdot C) \\
&\omega''_m{}^2 = \omega_0{}^2 / [(m^2 \pi^2 l_c{}^2 / 9a^2 + 1) \cdot (1 + 2\omega_0{}^2 \cdot M_2 \cdot C)
\end{aligned}\right\} \quad (97)$$

The solution (97) is by no means complete. We should like to recall that $M_1 \cdot C$ and $M_2 \cdot C$ are related to resonance frequencies resulting from modifications of the basis frequency. The solution (97) provides 'beat frequencies' with slow motions through and back along the DNA strands, i.e. the whole system is not rigid. These 'beat frequencies' are related to the above values of $\tau_0$, $\tau_1$, $\tau_2$, $\tau_3$, but further numerical values in the intervals of relevance exist.

The propagation of the changes of the configuration of the DNA bases obeys in the lowest order the solution (47), but we have also add correction terms

$$\left.\begin{aligned}
&Q(x,t) = \sum_m b_0 \cdot Q_{m,0} \cdot \sin(\pi \cdot m \cdot x / a) \cdot \cos(\omega_m \cdot t) + \\
&\sum_m b_1 \cdot Q_{m,0} \cdot (\sin(\pi \cdot m \cdot x / 2a) + \cos(\pi \cdot m \cdot x / 2a)) \cdot \cos(\omega'_m \cdot t) + \\
&\quad \sum_m b_2 \cdot Q_{m,0} \cdot \sin(\pi \cdot m \cdot x / 3a) \cdot \cos(\omega''_m \cdot t)
\end{aligned}\right\} \quad (98)$$

The solutions according to equations (97 – 98) are not independent, since the values for ω, ω', ω'' have to agree. In some model calculations, we have obtained the numerical values: $c_0 = 0.89$, $c_1 = 0.09$ $c_2 = 0.02$, $b_0 = 0.87$, $b_1 = 0.11$ $c_2 = 0.02$.

As already mentioned, this equation also exhibits standing waves. By specific interactions at some DNA-sites and energy supply the double stranded DNA is opened in a wave-like fashion, e.g. via H bond of water or a protein.

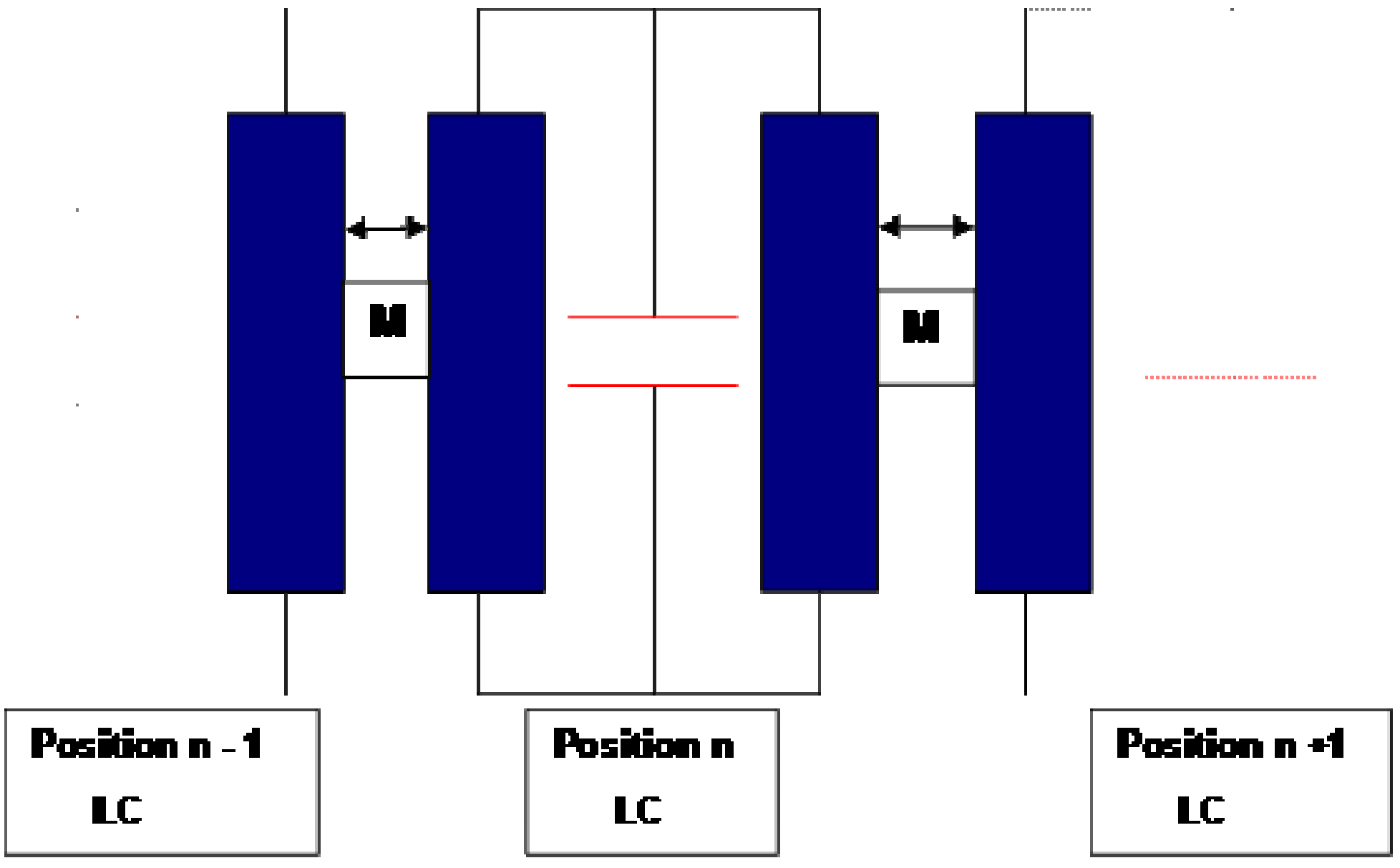

**Figure 13:** Periodic system of resonators with magnetic coupling M. The magnetic coupling can be replaced by an electrical coupling $C_{12}$. This leads to a similar type of a wave equation. Further terms result from couplings between non-adjacent neighbors.

## 3.2 Hydrolytic decay of ATP-Mg-Protein complexes

Figure 14 shows an ATP-Mg-protein complex, which one can find e.g. in the filaments of muscle cells. The energy of 0.5 eV is stored as a confined soliton (Davydov 1979). The soliton can certainly be considered as a standing wave of a system of oscillators with electrical coupling. The presence of $Ca^{2+}$ ions, water, tropomyosin, and troponin lead to the hydrolysis of ATP, and the coupling condition for standing waves, i.e. a node at the endpoints, breaks down. The stored energy escapes to become available for some other biomolecules, such as mechanical work of muscles, the muscle of the heart (very important) or synthesis of proteins, DNA replication and transcription, etc. It appears not to be probable that the stored soliton energy has very low frequencies. However, if $Ca^{2+}$ ions show a 7-days-period due to the solar magnetic field, then it becomes evident that ATP hydrolysis also occurs with a 7-days-period (Ulmer et al 2995, Ulmer 2002). The effect of the geomagnetic field leads to comparably fast oscillations (ca. 1 minute). The fast and slow oscillations occur simultaneously. Figure 14 presents the hydrolytic process in a schematic way. However, considering the magnetic interaction this process is rather complicated. According to the results in section 2.10.2 we have to account for the direction of the geomagnetic field and for the perpendicular component of the solar magnetic field. The contribution of the latter field is rather small, it leads mainly to an overall effect and can be compared with a bee-hive, which carries out an extremely slow pendular movement, whereas the bees are rather fast moving in various directions. The effective mass of the Mg has about magnitude as Ca, which implies nearly the same period (circaseptan) as Ca. The whole complex according to Figure 14 exhibits many degrees of freedom. One of them is the circular motion of the total complex in the geomagnetic field, which provides a circasemi-septan period, but this may be different in dependence of the mass number of the protein, i.e. the period could also be both longer and shorter. Further degrees are the large number of local oscillations of the positive and negative constituents of the phosphate groups and of the charged protein sites. These rotational oscillations are rather fast (order: some seconds). So the whole situation can be incorporated by coupled circuits with 4 or more resonators as previously shown. When we have a look to Figures 7a – 7d we can verify that the whole motion consisting of many degrees of freedom provides 'bear-frequencies' from one day up to 30 days. The components of circasemi-septan and circaseptan will become support by enhancement, which leads to an additional stabilization.

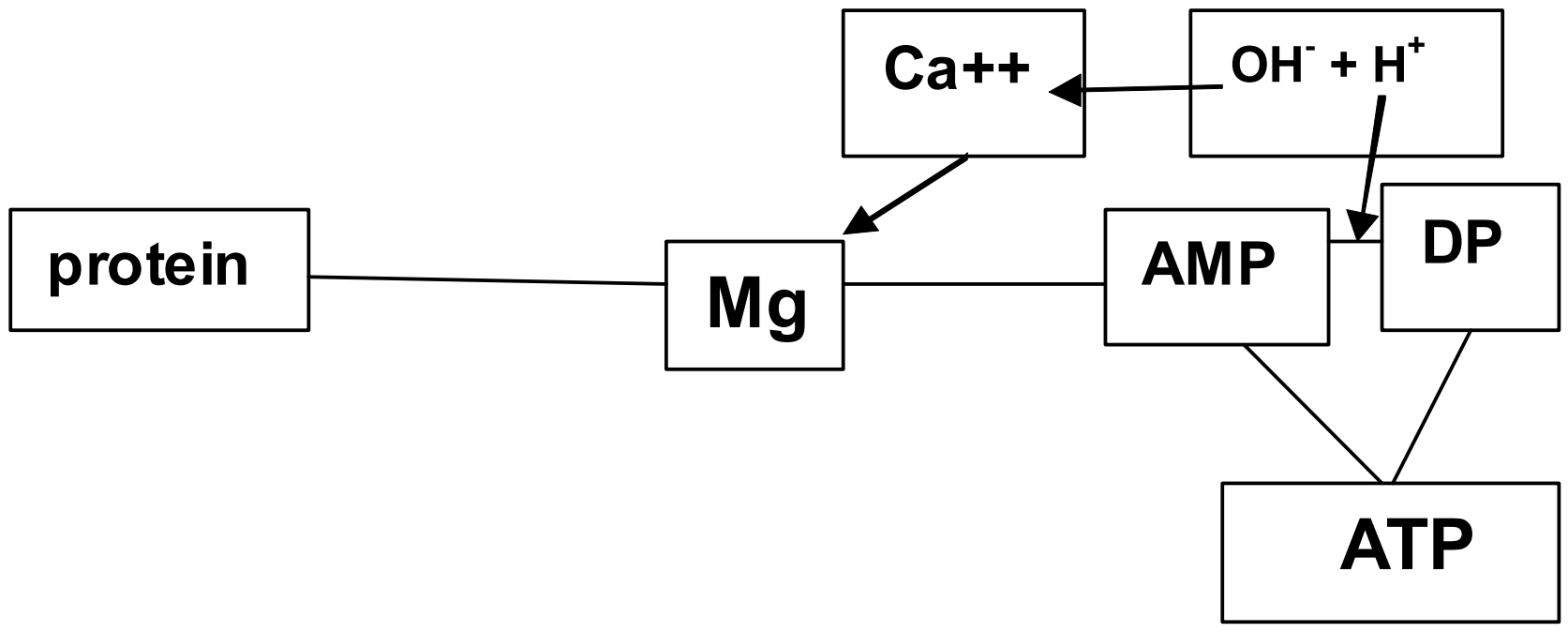

**Figure 14:** Mg-ATP-Protein complex (schematic representation) and hydroxylation by $Ca^{++}$ + $H_2O$.

The decay of ATP with release of a diphosphate group (DP in Figure 14) is founded by an H bond, i.e. one proton of the water molecule has to fulfill a tunnel process to reach an oxygen of ATP; therefore the symmetric potential for quantum mechanical tunneling is applicable (Figure 11). Due to the rotational oscillations induced by the geomagnetic field the overall motion is rather complex. Thus we have, at least, to be aware of spin-orbit and spin-spin coupling, which enhance the probability of quantum mechanical tunneling.

Due to the geomagnetic interaction the stored energy of 0.5 eV in the protein has to be a standing circular wave, which confirms the assumptions of Davydov based on the nonlinear Schrödinger equation (cited above).

## 4. Some conclusions

This section intends to develop a synthesis of the presented results with findings of some other authors.

At first, we remember the findings of Prigogine et al (1997). According to these authors nonlinear reaction-diffusion equations with feedback and feed-sideward couplings represent systems far from thermal equilibrium and show additionally the property of chemical/biological clocks (the Brusselator or the Oregonator are specific examples, see also Ulmer (2002)). The influences of magnetic fields (geomagnetic and solar magnetic) have not been accounted for by Prigogine et al. These influences may act in a cooperative way, i.e. they can lead to an enhancement or – if the magnetic resonances do not coincide with chemical clocks – to an inhibitory action. In every case, feedback and freed-sideward couplings of chemical reactions between specific molecules find their foundations in quantum mechanical resonance denominators analyzed in this communication. The term schemes and transition probabilities of two molecules have to be very similar to yield a chemical affinity of them. The resonance dominators induced by magnetic interactions lead to further splitting of the term schemes (fine structure splitting), and due to spin-orbit and spin-spin couplings intersystem crossing will be made possible.

Further important aspects are findings of Evstafyev (2009), Yoshii et al (2009), and Brown (1960, 1976) with regard to light-dependent influence of cryptochrome and magnetosensitivity of circadian clock and the influence of solar activity to molecular processes on earth. The communication of Yoshii et al is partially based on results of Brown (cited above). Thus a very important ‘Zeitgeber’ on earth is the light-dark cycle. This process might have been the origin of the circadian in evolution biology, but in course of the evolution the rhythm has become independent as shown in various experiments, i.e. it behaves as an intrinsic reaction process of specific biomolecules. From the viewpoint of term schemes it is evident that cryptochome and cytochrome molecules containing FAD, NADP, NADPH and interacting with tryptophan and its derivatives melatonin and serotonin are readily excited by external light in the day rhythm to induce excited singlet and triplet states (see term scheme 9). Since visible light is absorbed in the skin, transport mechanisms leading to long-range interactions are required to affect further molecular processes. The discussed magnetic properties now become rather significant, since only intersystem crossing can lead to transport of energy. In long molecular chains pure singlet excitations/transitions imply only local excitons which are damped by scatter of light and heat production. An important transport mechanism is represented by a chain of H bonds in water (Figures 15a, 15b), since water molecules are always present in cellular tissue.

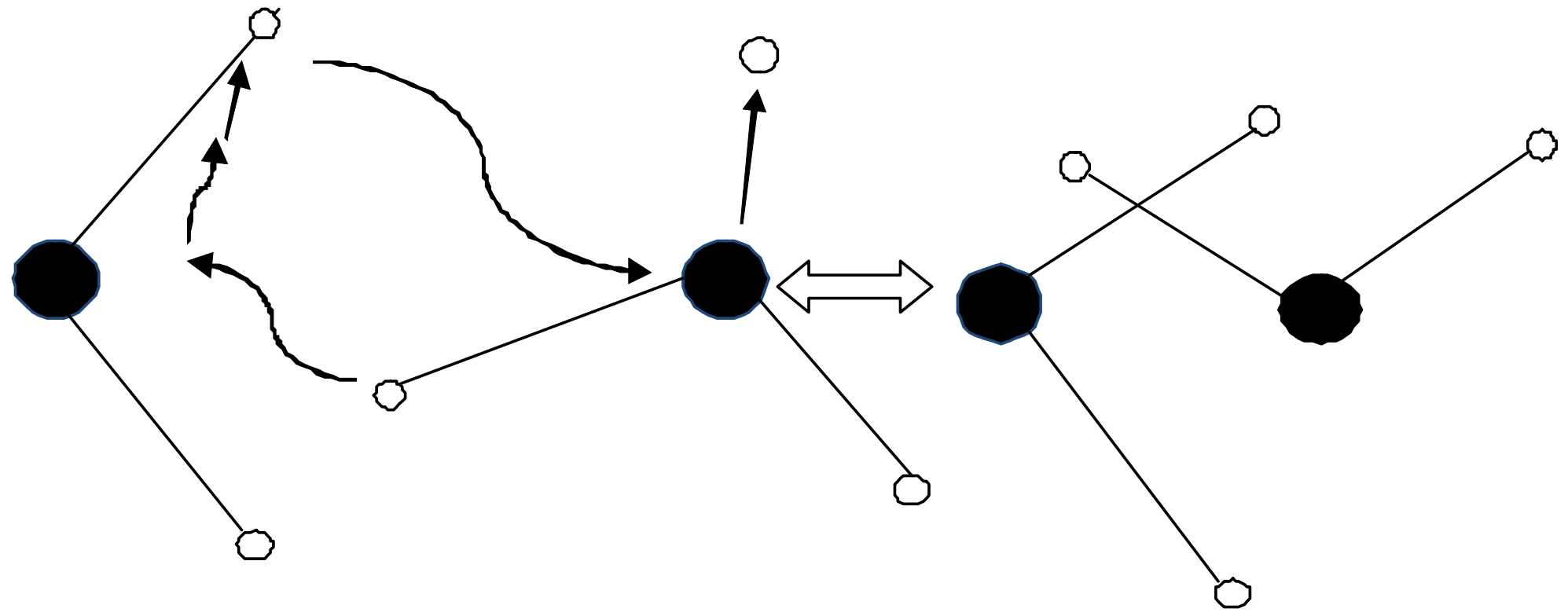

**Figure 15a:** Proton exchange between two water molecules (left: initial state, right: final state).

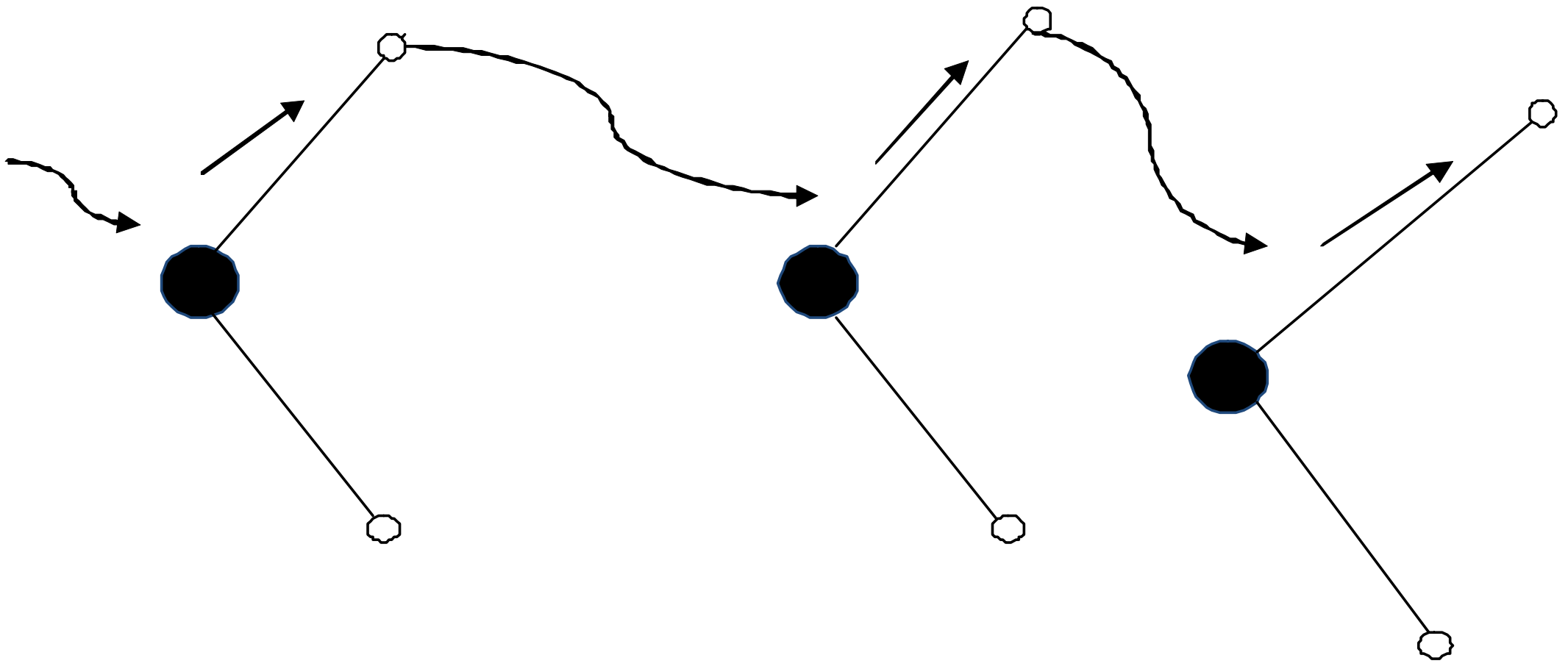

**Figure 15b:** H bond with exchange of protons in a chain of water molecular: possible mechanism of energy transfer in cells and intercellular medium.

The results obtained by (cited above) mainly deal with the role of spin-orbit coupling, singlet-triplet transitions and spin waves in systems far from thermal equilibrium, where the $k_B \cdot T$ rule is violated due to phase transitions and critical points. A particular stress are frequency bands from 10 Hz (origin: earth, circadian rhythm, Engelmann et al 1996)) and the MHz domain (Evstafyev), which has solar origin. The associated energy lies below the energy $k_B \cdot T$ in thermal equilibrium and can only affect and synchronize those processes connected with magnetic properties at critical points to prevent thermal equilibrium, which would be death of living systems. Since the connection of electromagnetic properties to biorhythms is a proven fact, the viewpoint of the magnitude of the energy can certainly not be valid. As a final example we present own measurement results of the ATP metabolism (growth and stationary phase) obtained via 31-P NMR spectroscopy in tumor spheroids. The measurement conditions have been published elsewhere. These measurements show, in addition to rather fast processes, typical cycles, of which the circaseptan, circasemi-septen are very prominent. The circadian period is also present, but it appears not to play the dominant role. The measurements have been carried out in complete darkness; therefore it is ensured that the light-dark rhythm can be excluded. There have been put forward new measurement methods to quantify ATP in cellular processes, which avoid the strong magnetic fields of NMR spectroscopy (Lundin et al 1975, Masson et al 2008, Weber et al 2010).

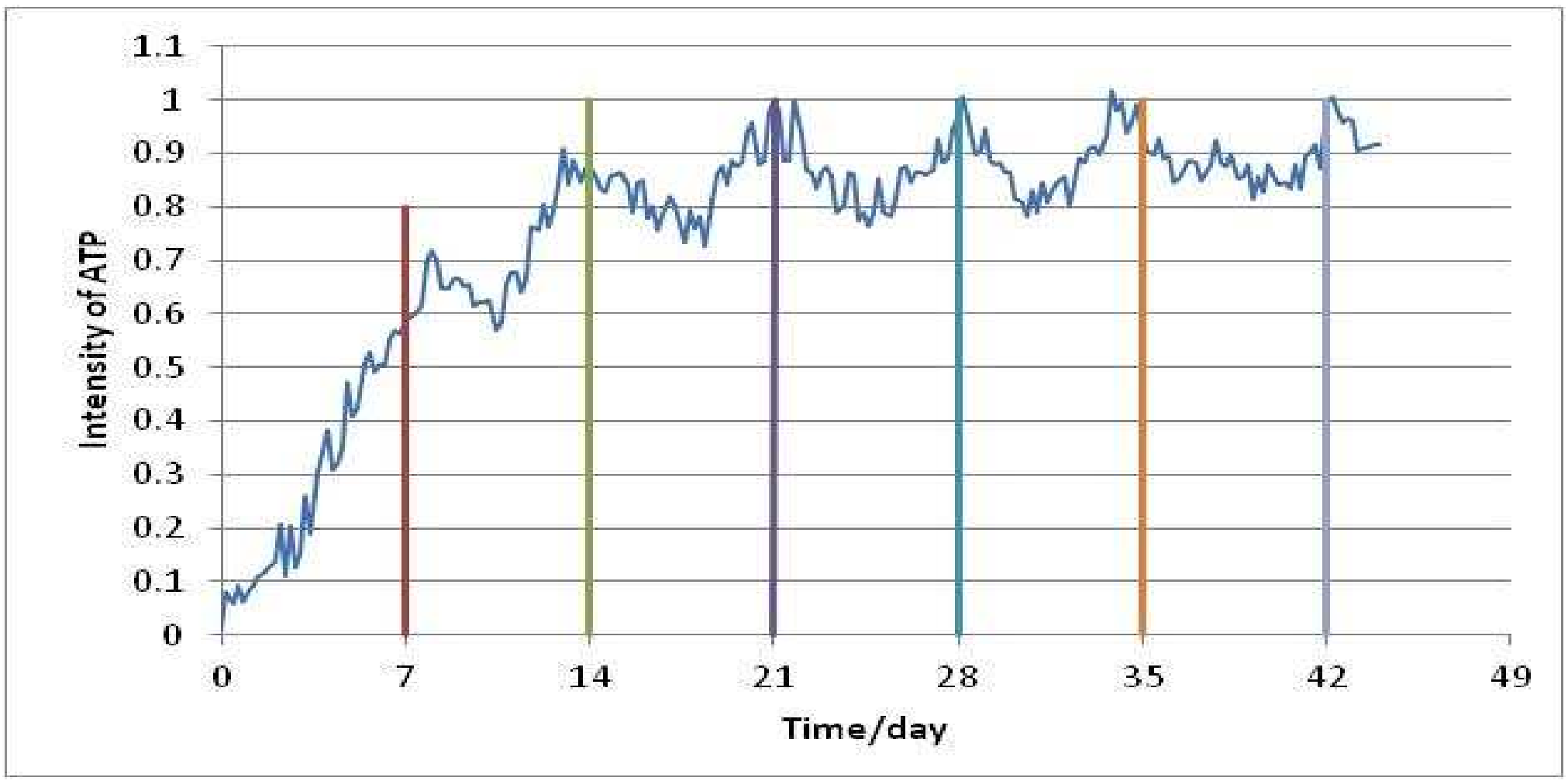

**Figure 16:** ATP concentration in tumor spheroids (growth and plateau phase) of C3H mammalian cells.

These methods appear to be much closer to the cellular physiology and may serve to yield a better optimization of cancer radiotherapy (Ulmer 2002).

## References

M. Abramowitz, and I. Stegun (1970) Handbook of Mathematical Functions with Formulas, Graphs and Mathematical Tables, Natural Bureau of Standards

G. M. Barenboim, A. N. Domanski and K. K. Turerov (1969) Luminescence of biopolymers and cells Plenum Press, London – New York

J. B. Birks (1970) Photo physics of Aromatic Molecules Wiley-Interscience, London

F. A. Brown (1960) Cold Spring Harbor Symposium Quantum Biology 25 57 – 71

F. A. Brown (1976) Bio Systems 8 67 – 81

D. K. Campbell and M. Peyrar (1983) Physics 9D, 3

F.L. Carter (1981) In Molecular Electronics Devices, Marcel Dekker, New York

A.S. Davydov (1979) International Journal of Quantum Chemistry, 16, 5

W. Engelmann, W. Hellrung and A. Johnson (1996) Bioelectromagnetics 17 100 – 100

V. K. Evastafev (2009) The Open Biology Journal 2 38 – 41

K. Fukui and H. Fujimoto (1969) Bull. Chem. Soc. Japan 42 3399

F. Halberg, M. Engeli, C. Hamburger and D. Hillman (1965) Acta endocrinol (Kbh) 50 (Suppl 103) 5 - 54

F. Halberg and G. Cornélissen (1991) Chronobiologia 18 114 – 120

H. F. Hameka (1962a) Journal of Chemical Physics 37 328

H. F. Hameka (1962b) Journal of Chemical Physics 37 2209

H. F. Hameka (1965) Advanced Quantum Chemistry Addison-Weyley Publishing Company Inc., Massachusetts

H. Hartman and W. Stürmer (1950) Z. Naturforschung 6a, 751

H. Hartmann (1964) Theorie der chemischen Bindung Springer Heidelberg

T. A. Hoffman and J. Ladik (1962) Cancer Research 21 474

A. O. von Lilienfeld and A. Tkatchenko (2010) J. Chem. Phys. 133 084104

A. Lundin and A. Thore (1975) Analyt. Biochemistry 66 47 – 63

R. Mason (1958) Nature 181 820 – 827

J. F. Masson, C. Kranz, B. Mizaikoff and E. B. Gauda (2008) Analyt. Chem. 89 3991 – 3998

A. Pérez, M. Tuckerman, H. Hjalmarson and O. A. von Lilienfeld (2010) JACS 132 11510 – 11515

F. A. Popp, G. Becker, H.König and W. Peschka (1979) Electromagnetic Bio-Information Urban & Schwarzenberg München/Wien/Baltimore

I. Prigogine and I. Stengers (1997) The End of Certainty, Time, Chaos and the New Laws of Nature The Free Press, New York

A. Pullman and B. Pullman (1962) Nature 190 228

D. K. Randhawa, L. M. Kaur and M. L. Singh (2011) Int. Journ. Computer Applications **17** 8 – 12

H. Schweiger, S. Berger, H. Kretschmer, A. Mörler, E. Halberg, R. Sothern and F. Halberg (1986) Proc. Nat. Acad. Science USA 83 8619 - 8623

P. Schuster, G. Zundel and C. Sandorfy (1976) The Hydrogen Bond – Recent Deveopments in Theory and Experiments North-Holland Publishing Company, Amsterdam and New York

S. S. Sung (1977) International Journal Quantum Chemistry XII, Supplement I 387

W. Ulmer and H. Hartmann (1978) Nuovo Cimento A 47 59 – 78

W. Ulmer (1980a) Theoretica Chimica Acta, 55, 179 – 201

W. Ulmer (1980b) Theoretica Chimica Acta 56, 133 – 148

W. Ulmer (1979) Z. Naturforschung 34c, 658 – 669

W. Ulmer (1981) Int. Journ. Quantum Chemistry 19, 337 – 359

W. Ulmer (1983) Int. Journ. Quantum Chemistry 23 1931 – 1945

W. Ulmer, G. Cornélissen and F. Halberg (1995) In vivo 9 363 – 374

W. Ulmer (2002) In vivo 16 31 – 36

C. Weber, E. Gauda, E. Hecht, B. Mizaikoff and C. Kranz (2010) IFMBE Proceedings Springer Heidelberg 25/8 351 – 354

T. Yoshii, M. Ahmad and C. H. Helfrich-Förster (2009) Journal Plosbiology 7 e1000086